\documentclass[10pt,journal,onecolumn]{IEEEtran}

\usepackage{amsmath,amsfonts,bm}

\def\ceil#1{\lceil #1 \rceil}

\def\1{\bm{1}}

\def\vmu{{\bm{\mu}}}

\def\vw{{\bm{w}}}

\def\vz{{\bm{z}}}

\def\mA{{\bm{A}}}

\def\mN{{\bm{N}}}

\def\mX{{\bm{X}}}
\def\mY{{\bm{Y}}}

\def\mSigma{{\bm{\Sigma}}}

\DeclareMathAlphabet{\mathsfit}{\encodingdefault}{\sfdefault}{m}{sl}
\SetMathAlphabet{\mathsfit}{bold}{\encodingdefault}{\sfdefault}{bx}{n}

\newcommand{\E}{\mathbb{E}}

\DeclareMathOperator*{\Cov}{Cov}

\DeclareMathOperator*{\argmax}{arg\,max}

\DeclareMathOperator*{\argmin}{arg\,min}
\DeclareMathOperator*{\arginf}{arg\,inf}

\DeclareMathOperator*{\tr}{tr}

\usepackage[colorlinks=true,citecolor = black]{hyperref}%
\usepackage{url}            
\usepackage{booktabs}       
\usepackage{amsfonts}       
\usepackage{microtype}      

\usepackage{cite,calc}
\usepackage{algorithmic}
\usepackage{graphicx}
\usepackage{makecell}

\usepackage{amssymb}

\usepackage{xcolor}

\newtheorem{theorem}{Theorem}
\newtheorem{lemma}{Lemma}
\newtheorem{assumption}{Assumption}

\newtheorem{proposition}{Proposition}

\newtheorem{corollary}{Corollary}

\newtheorem{definition}{Definition}
\newtheorem{example}{Example}

\newtheorem{remark}{Remark}

\newcommand{\nn}{\notag}
\newcommand{\diff}{\mathrm{d}}
\DeclareMathOperator*{\TV}{TV}
\DeclareMathOperator*{\gen}{gen}
\newcommand{\genb}{\overline\gen}

\definecolor{yuheng}{rgb}{0.133, 0.6, 0.329}
\definecolor{gholamali}{rgb}{0.204, 0.596, 0.859}

\title{Information-theoretic Characterizations of Generalization Error for the Gibbs Algorithm}

\author{Gholamali Aminian$^*$,~\IEEEmembership{Member,~IEEE,}
        Yuheng Bu$^*$,~\IEEEmembership{Member,~IEEE,}
        Laura Toni,~\IEEEmembership{Senior~Member,~IEEE,}
        Miguel R. D. Rodrigues,~\IEEEmembership{Senior~Member,~IEEE,}
        Gregory W. Wornell,~\IEEEmembership{Fellow, IEEE}
\thanks{$^*$ Equal Contribution.}
\thanks{This work was presented in part at the Conference on Advances in Neural Information Processing Systems (NeurIPS), Virtual Conference, 2021 \cite{aminian2021exact}.
}
\thanks{G. Aminian is with the Alan Turing Institute, London, UK (Email: gaminian@turing.ac.uk).}
\thanks{Y. Bu was with the Department of Electrical Engineering and Computer Science, Massachusetts Institute of Technology, Cambridge, MA 02139. He is now with the Department of Electrical and Computer Engineering, University of Florida, Gainesville, FL 32611 (Email: buyuheng@ufl.edu).}
\thanks{L. Toni and M. Rodrigues are with the Electronic and Electrical Engineering Department at University College London, UK (Email:  l.toni, m.rodrigues@ucl.ac.uk).}
\thanks{G. W. Wornell is with the Department of Electrical Engineering and Computer Science, Massachusetts Institute of Technology, Cambridge, MA 02139 (Email: gww@mit.edu).}
\thanks{This work has been supported in part by the MIT-IBM Watson AI Lab under Agreement No. W1771646, AFRL under Cooperative Agreement No.~FA8750-19-2-1000, NSF under Grant No. CCF-1816209, the UKRI Prosperity Partnership Scheme (FAIR) under the EPSRC Grant EP/V056883/1  and the Alan Turing Institute. Gholamali Aminian is further supported in part by the Royal Society Newton International Fellowship, project no. NIF\textbackslash R1 \textbackslash192656. }
}

\begin{document}
\maketitle


\begin{abstract}
Various approaches have been developed to upper bound the generalization error of a supervised learning algorithm. However, existing bounds are often loose and even vacuous when evaluated in practice. As a result, they may fail to characterize the exact generalization ability of a learning algorithm.
Our main contributions are exact characterizations of the expected generalization error of the well-known Gibbs algorithm (a.k.a. Gibbs posterior) using different information measures, in particular, the symmetrized KL information between the input training samples and the output hypothesis. Our result can be applied to tighten existing expected generalization error and PAC-Bayesian bounds. Our information-theoretic approach is versatile, as it also characterizes the generalization error of the Gibbs algorithm with a data-dependent regularizer and that of the Gibbs algorithm in the asymptotic regime, where it converges to the standard empirical risk minimization algorithm. Of particular relevance, our results highlight the role the symmetrized KL information plays in controlling the generalization error of the Gibbs algorithm.
\end{abstract}




\section{Introduction}\label{Sec:Introduction}

\IEEEPARstart{U}{nderstanding} the generalization behavior of a learning algorithm is one of the most important challenges in statistical learning theory. 
Various approaches have been developed~\cite{rodrigues2021information}, including VC dimension-based bounds~\cite{vapnik1999overview}, algorithmic stability-based bounds ~\cite{bousquet2002stability}, algorithmic robustness-based bounds ~\cite{xu2012robustness}, PAC-Bayesian bounds~\cite{mcallester2003pac}, and recently information-theoretic bounds~\cite{xu2017information}. 

However, upper bounds on the generalization error cannot entirely capture the generalization ability of a learning algorithm. One apparent reason is the tightness issue, and some upper bounds \cite{anthony2009neural} can be far away from the true generalization error or even vacuous when evaluated in practice. More importantly, existing upper bounds do not fully characterize all the aspects that could influence the generalization error of a supervised learning problem. For example, VC dimension-based bounds depend only on the hypothesis class, and algorithmic stability-based bounds only exploit the properties of the learning algorithm. As a consequence, both methods fail to capture the fact that the generalization error depends strongly on the interplay between the hypothesis class, learning algorithm, and the underlying data-generating distribution, as discussed in \cite{zhang2021understanding,xu2017information}.
This paper adopts an information-theoretic approach to overcome the above limitations by deriving  \emph{exact}  characterizations of the generalization error for a specific supervised learning algorithm, namely the Gibbs algorithm.



\subsection{Problem Formulation}

Let $S = \{Z_i\}_{i=1}^n \in \mathcal{S}$ be the training set, where each $Z_i=\{X_i,Y_i\}$ is defined on the same alphabet $\mathcal{Z}$. Note that $Z_i$ is not required to be i.i.d generated from the same data-generating distribution $P_Z$, and we denote the joint distribution of all the training samples as $P_S$.
We also denote the hypotheses by $w \in \mathcal{W}$, where $\mathcal{W}$ is a hypothesis class. The performance of the hypothesis is measured by a non-negative loss function $\ell:\mathcal{W} \times \mathcal{Z}  \to \mathbb{R}_0^+$, and we define the empirical and population risks associated with a given hypothesis $w$ via 
\begin{align*}
    L_\mathrm{e}(w,s)\triangleq\frac{1}{n}\sum_{i=1}^n \ell(w,z_i)
    \intertext{and}
    L_\mathrm{p}(w,P_S)\triangleq  \mathbb{E}_{P_S}[L_\mathrm{e}(w,S)],
\end{align*}
respectively. A learning algorithm can be modeled as a randomized mapping from the training set $S$ onto an hypothesis $W\in\mathcal{W}$  according to the conditional distribution $P_{W|S}$. Thus, the expected generalization error quantifying the degree of over-fitting can be written as
\begin{equation}\label{Eq: expected GE}
\genb(P_{W|S},P_S)\triangleq\mathbb{E}_{P_{W,S}}[ L_\mathrm{p}(W,P_S)-L_\mathrm{e}(W,S)],
\end{equation}
where the expectation is taken over the joint distribution $P_{W,S} =  P_{W|S}\otimes P_S$. 


In this paper we focus on the generalization error of the Gibbs algorithm (a.k.a. Gibbs posterior~\cite{catoni2007pac}). The $(\gamma,\pi(w),f(w,s))$-Gibbs distribution, which was first proposed by \cite{gibbs1902elementary} in statistical mechanics and further investigated by \cite{jaynes1957information} in information theory, is defined as:
\begin{equation}\label{Eq: Gibbs Solution}
    P_{{W}|S}^\gamma (w|s) \triangleq \frac{\pi({w})\, e^{-\gamma f(w,s)}}{V_f(s,\gamma)},\quad \gamma\ge 0,
\end{equation}
where $\gamma$ is the inverse temperature, $\pi(w)$ is an arbitrary prior distribution of $W$, $f(w,s)$ is energy function, and 
\begin{equation*}
V_f(s,\gamma) \triangleq \int \pi(w) e^{-\gamma f(w,s)} \diff w
\end{equation*}
is the partition function. 

In this paper, we will characterize the generalization behaviour of the Gibbs algorithm via various information measure. In particular, if $P$ and $Q$ are probability measures over space $\mathcal{X}$, and $P$ is absolutely continuous with respect to $Q$, the Kullback-Leibler (KL) divergence between $P$ and $Q$ is given by
\begin{equation*}
   D(P\|Q)\triangleq\int_\mathcal{X}\log\left(\frac{dP}{dQ}\right) dP. 
\end{equation*}
If $Q$ is also absolutely continuous with respect to $P$, then the symmetrized KL divergence (also referred to as Jeffrey's divergence \cite{jeffreys1946invariant}) is 
\begin{equation}
D_{\mathrm{SKL}}(P\|Q)\triangleq D(P \| Q) + D(Q\|P).
\end{equation}
The $\alpha$-R\'enyi divergence between $P$ and $Q$ is given by \begin{equation*}
R_\alpha(P\|Q)=\frac{1}{\alpha-1}\log(\int_{\mathcal{X}}(dP)^\alpha(dQ)^{1-\alpha}),\quad \alpha\ge 0.
\end{equation*}
The mutual information between two random variables $X$ and $Y$ is defined as the KL divergence between the joint distribution and product-of-marginal
distribution 
\begin{equation*}
I(X;Y)\triangleq D(P_{X,Y}\|P_X\otimes P_{Y}),
\end{equation*}
or equivalently, the conditional KL divergence between $P_{Y|X}$ and $P_Y$ over $P_X$, $D(P_{Y|X} \| P_Y|P_{X})\triangleq\int_\mathcal{X}D(P_{Y|X=x} \| P_Y)dP_{X}(x)$.
By swapping the role of $P_{X,Y}$ and $P_X\otimes P_{Y}$ in mutual information, we get the lautum information introduced by \cite{palomar2008lautum}, $L(X;Y)\triangleq D(P_X\otimes P_{Y}\| P_{X,Y})$.

Finally, the symmetrized KL information between $X$ and $Y$ is given by \cite{aminian2015capacity}:
\begin{equation}
   I_{\mathrm{SKL}}(X;Y)\triangleq D_{\mathrm{SKL}}(P_{X,Y}\|P_X\otimes P_Y)= I(X;Y)+ L(X;Y). 
\end{equation}
The conditional mutual information between two random variables $X$ and $Y$ conditioned on $Z$ is the KL divergence between $P_{X,Y|Z}$ and $P_{X|Z} \otimes P_{Y|Z}$ averaged over $P_{Z}$, $$I(X;Y|Z)\triangleq\int_\mathcal{Z}D(P_{X,Y|Z=z} \| P_{Y|Z=z} \otimes P_{X|Z=z}) dP_{Z}(z).$$ Similarly, we can also define the conditional lautum information $L(X;Y|Z)$, and the conditional symmetrized KL information is given by
\begin{align}
   &I_{\mathrm{SKL}}(X;Y|Z)\triangleq
   I(X;Y|Z)+ L(X;Y|Z). 
\end{align}

Throughout the paper, upper-case letters denote random variables (e.g., $Z$), lower-case letters denote the realizations of random variables (e.g., $z$), and calligraphic letters denote sets (e.g., $\mathcal{Z}$). 
All the logarithms are the natural ones, and all the information measure units are nats. $\mathcal{N}(\mu,\Sigma)$ denotes the Gaussian distribution with mean $\mu$ and covariance matrix $\Sigma$.

\subsection{Contributions}
The core contribution of this paper (see Theorem \ref{Theorem: Gibbs Result}) is an \emph{exact} characterization of the expected generalization error for the Gibbs algorithm in terms of the symmetrized KL information between the input training samples $S$ and the output hypothesis $W$, as follows:
\begin{equation*}
    \genb(P_{{W}|S}^\gamma,P_S) =  \frac{I_{\mathrm{SKL}}(W;S)}{\gamma}.
\end{equation*}
We also discuss some general properties of the symmetrized KL information, which could be used to prove the non-negativity and concavity of the expected generalization error for the Gibbs algorithm.

{\color{black}In addition, we provide exact characterizations of the expected generalization error using other information measures, including symmetrized KL divergence, conditional symmetrized KL information, and replace-one symmetrized KL divergence. These results highlight the fundamental role of symmetrized KL information (divergence) in learning theory that does not appear to have been recognized before. } 

Building upon these results, we further expand our contributions in various directions:
\begin{itemize}
   {\color{black} \item In Section \ref{sec:upper_bound},  we tighten existing expected generalization error bound (see Theorem~\ref{Theorem: Sub Gaussian extension}, Theorem~\ref{Theorem: param upper on Gibbs based on KL}, Theorem~\ref{Theorem: param via conditional SYI} and Theorem~\ref{Theorem: param upper on Gibbs using stability}) by combining our exact characterizations of expected generalization error with the existing bounding techniques. 
    \item In Section \ref{sec:upper_bound}, we also tighten the PAC-Bayesian bound (see Theorem~\ref{Theorem: PAC Baysian result}) for Gibbs algorithm under i.i.d and sub-Gaussian assumptions using symmetrized KL divergence.}
    \item In Section \ref{sec:asymptotic} (Proposition~\ref{prop:single_well} and \ref{prop:multiple_well}), we show how to use our method to characterize the asymptotic behavior of the generalization error for Gibbs algorithm under large inverse temperature limit $\gamma\to \infty$, where Gibbs algorithm converges to the empirical risk minimization algorithm. Note that existing bounds, such as \cite{kuzborskij2019distribution,xu2017information,raginsky2016information}, become vacuous in this regime. 
    \item In Section \ref{sec:regularization}, we characterize the generalization error of the Gibbs algorithm with data-dependent regularizer using symmetrized KL information, which provides some insights on how to reduce the generalization error using regularization.
\end{itemize}

Some of these contributions have been presented in part in \cite{aminian2021exact}. However, this paper generalize \cite{aminian2021exact} by deriving new exact characterizations. We also utilise these exact characterizations to derive upper bounds that are more tight.
\subsection{Motivations for the Gibbs Algorithm}
As we discuss below, the choice of the Gibbs algorithm is not arbitrary since it arises naturally in many different applications and is sufficiently general to model many learning algorithms used in practice:

\textbf{Empirical Risk Minimization:}  The $(\gamma,\pi(w),L_\mathrm{e}(w,s))$-Gibbs algorithm can be viewed as a randomized version of empirical risk minimization (ERM) if we specify the energy function $f(w,s) = L_\mathrm{e}(w,s)$. As the inverse temperature $\gamma \to \infty$, the prior distribution $\pi(w)$ becomes negligible, and hypothesis generated by the Gibbs algorithm converges to the hypothesis corresponding to standard ERM.

\textbf{Information Risk Minimization:} The Gibbs algorithm also arises when conditional KL-divergence is used as a regularizer to penalize over-fitting in the information risk minimization framework. In particular, it is shown in \cite{xu2017information,zhang2006information,zhang2006E} that the solution to the following regularized ERM problem
\begin{align}\label{Eq: optimization problem}
&P^{\star}_{W|S} \notag\\
&\quad =\arginf_{P_{W|S}}\left( \mathbb{E}_{P_{W,S}}[L_\mathrm{e}(W,S)]+\frac{1}{\gamma} D(P_{W|S}\|\pi(W)|P_S)\right)
\end{align}
corresponds to the $(\gamma,\pi(w),L_\mathrm{e}(w,s))$-Gibbs distribution. The inverse temperature $\gamma$ controls the regularization term and balances between over-fitting and generalization.

\textbf{PAC-Bayesian Bound:} The following upper bound on the population risk from \cite{thiemann2017strongly} holds  with probability at least $1-\delta$ for $0<\delta<1$, and $0<\lambda<2$ under distribution $P_S$,
\begin{align}\label{Eq: PAC Thiemann bound}
&\mathbb{E}_{P_{W|S=s}}[L_\mathrm{p}(W,P_S)]\notag\\
&\quad \leq \frac{\mathbb{E}_{P_{W|S=s}}[L_\mathrm{e}(W,s)]}{1-\frac{\lambda}{2}}+\frac{D(P_{W|S=s}\| \pi(W))+\log(\frac{2\sqrt{n}}{\delta})}{\lambda(1-\frac{\lambda}{2})n }.
\end{align}
If we fix $\lambda$, $\pi(w)$ and optimize over $P_{W|S=s}$, the distribution that minimizes the PAC-Bayes bound in \eqref{Eq: PAC Thiemann bound} is the $(n\lambda,\pi(w),L_\mathrm{e}(w,s))$-Gibbs distribution. Similar bounds are proposed in \cite[Theorem~1.2.1]{catoni2007pac} and \cite[Lemma~10]{van2014pac}, where optimizing over the posterior distribution would result in a Gibbs distribution.

 \textbf{SGLD Algorithm:} 
The continuous-time Langevin diffusion is described by the following stochastic differential equation of a random process $W(t)$:
\begin{equation}\label{Eq: Langevin diffusion}
    \diff W(t)=-\nabla L_\mathrm{e}(W(t),s)\, \diff t + \sqrt{\frac{2}{\gamma}}\, \diff B(t), \quad t \ge 0,
\end{equation}
where $B(t)$ is the standard Brownian motion. Under some conditions on the loss function $\ell(w,z)$, \cite{chiang1987diffusion,markowich2000trend} shows that in the continuous-time Langevin diffusion, the stationary distribution of hypothesis $W$ is the Gibbs distribution. 

The Stochastic Gradient Langevin Dynamics (SGLD) can be viewed as the discrete version of the continuous-time Langevin diffusion, and it is defined as follows:
\begin{equation}
    W_{k+1}=W_k-\beta\,\nabla L_\mathrm{e}(W_k,s)+\sqrt{\frac{2\beta}{\gamma}}\,\zeta_k, \quad k=0,1,\cdots,
\end{equation}
where $\zeta_k$ is a standard Gaussian random vector  and $\beta>0$ is the step size. In \cite{raginsky2017non}, it is proved that under some conditions the on loss function, the conditional distribution $P_{W_k|S}$ induced by SGLD algorithm is close to the $(\gamma,\pi(W_0),L_\mathrm{e}(w_k,s))$-Gibbs distribution in 2-Wasserstein distance for sufficiently large $k$.

\subsection{Other Related Work}\label{subsec:related_work}
\textbf{Information-theoretic generalization error bounds:} Recently, \cite{russo2019much,xu2017information} propose to use the mutual information between the input training set and the output hypothesis to upper bound the expected generalization error. However, those bounds are known not to be tight, and multiple approaches have been proposed to tighten the mutual information-based bound. \cite{bu2020tightening} provides tighter bounds by considering the individual sample mutual information, \cite{asadi2018chaining,asadi2020chaining} propose using chaining mutual information,  and \cite{steinke2020reasoning,hafez2020conditioning,haghifam2020sharpened} advocate the conditioning and processing techniques. 
Information-theoretic generalization error bounds using other information quantities are also studied, such as, $f$-divergence~\cite{jiao2017dependence}, $\alpha$-R\'enyi divergence and maximal leakage~\cite{issa2019strengthened,esposito2019generalization}, Jensen-Shannon divergence~\cite{aminian2020jensen,aminian2022learning} and Wasserstein distance~\cite{aminian2022tighter,lopez2018generalization,wang2019information,rodriguez2021tighter}. Using rate-distortion theory, \cite{masiha2021learning,bu2020information,bu2021population} provide information-theoretic generalization error upper bounds for model misspecification and model compression. The information-theoretic approaches are also applied for the generalization error of other scenarios including, semi-supervised learning~\cite{aminian2022information,Haiyun2022Gibbs}, transfer learning~\cite{bu2022characterizing} and meta learning~\cite{jose2021information}.





\textbf{PAC-Bayesian generalization error bounds:} First proposed by \cite{shawe1997pac,mcallester1999some,mcallester2003pac}, PAC-Bayesian analysis provides high probability bounds on the generalization error in terms of KL divergence between the data-dependent posterior induced by the learning algorithm and a data-free prior that can be chosen arbitrarily~\cite{alquier2021user}. There are multiple ways to generalize the standard PAC-Bayesian bounds, including using different information measures other than the KL divergence \cite{ohnishi2021novel,alquier2018simpler,begin2016pac,hellstrom2020generalization,aminian2021information} and considering data-dependent priors (prior depends on the training data) \cite{rivasplata2020pac,dziugaite2018data,catoni2007pac,banerjee2021information,dziugaite2018entropy,ambroladze2007tighter} or distribution-dependent priors (prior depends on data-generating distribution) \cite{catoni2003pac,parrado2012pac,lever2013tighter,rivasplata2018pac}. In \cite{germain2015risk}, a more general PAC-Bayesian framework is proposed, which provides a high probability bound on the convex function of the expected population and empirical risk with respect to the posterior distribution, whereas in \cite{germain2016pac} the connection between Bayesian inference and PAC-Bayesian theorem is explored by considering Gibbs posterior and negative log loss function.

\textbf{Generalization error of Gibbs algorithm:} Both information-theoretic and PAC-Bayesian approaches have been used to bound the generalization error of the Gibbs algorithm.  An information-theoretic upper bound with a convergence rate of $\mathcal{O}\left(\gamma/n\right)$ is provided in \cite{raginsky2016information} for the Gibbs algorithm with bounded loss function, and PAC-Bayesian bounds using a variational approximation of Gibbs posteriors are studied in \cite{alquier2016properties}. \cite[Appendix~D]{asadi2020chaining} provides an upper bound on the excess risk of the Gibbs algorithm under the sub-Gaussian assumption. \cite{kuzborskij2019distribution} focuses on the excess risk of the Gibbs algorithm, and a similar generalization bound with a rate of $\mathcal{O}\left(\gamma/n\right)$ is provided under the sub-Gaussian assumption. Although these bounds are tight in terms of the sample complexity $n$, they become vacuous when the inverse temperature $\gamma \to \infty$, hence are unable to capture the behavior of the ERM algorithm. The sensitivity of the expected empirical risk with respect to the Gibbs algorithm is studied in \cite{perlaza2022empirical}.

Our work differs from this body of research because we provide exact characterizations of the generalization error of the Gibbs algorithm in terms of different information measures.  Our work further leverages this characterization to tighten existing expected and PAC-Bayesian generalization error bounds in literature such as~\cite{kuzborskij2019distribution,lever2013tighter}.

\section{Generalization Error of the Gibbs Algorithm}\label{Sec:Main Results}

Our main result, which characterizes the exact expected generalization error of the Gibbs algorithm with prior distribution $\pi(w)$, is as follows:
\begin{theorem}\label{Theorem: Gibbs Result}
For $(\gamma,\pi(w),L_\mathrm{e}(w,s))$-Gibbs algorithm,
\begin{equation}\label{equ:Gibbsposterior1}
    P_{{W}|S}^\gamma (w|s) = \frac{\pi({w})\, e^{-\gamma L_\mathrm{e}(w,s)}}{V_{L_\mathrm{e}}(s,\gamma)},\quad \gamma>0,
\end{equation}
the expected generalization error is given by
\begin{equation}
    \genb(P_{{W}|S}^\gamma,P_S) =  \frac{I_{\mathrm{SKL}}(W;S)}{\gamma}.
\end{equation}
\end{theorem}
\begin{IEEEproof}[\textbf{Sketch of Proof}] It can be shown that the symmetrized KL information can be written as
\begin{equation}\label{eq:symetrized KL representation}
    I_{\mathrm{SKL}}(W;S)=\mathbb{E}_{P_{W,S}}[\log(P_{W|S}^\gamma)]-\mathbb{E}_{P_W \otimes P_S}[\log(P_{W|S}^\gamma)].
\end{equation}
Just like the generalization error, the above expression is the difference between 
the expectations of the same function evaluated under the joint distribution and the product-of-marginal distribution. Note that $P_{W,S}$ and $P_W \otimes P_S$ share the same marginal distribution, we have $\mathbb{E}_{P_{W,S}}[\log\pi(W)] = \mathbb{E}_{P_W}[\log\pi(W)]$, and 
$\mathbb{E}_{P_{W,S}}[\log V_{L_\mathrm{e}}(S,\gamma)] = \mathbb{E}_{P_S}[\log V_{L_\mathrm{e}}(S,\gamma)]$. Then,  combining \eqref{equ:Gibbsposterior1} with \eqref{eq:symetrized KL representation} completes the proof.  More details together with the full proof are provided in Appendix~\ref{app: Gibbs Algorithm}.
\end{IEEEproof}
To the best of our knowledge, this is the first exact characterization of the expected generalization error for the Gibbs algorithm. Note that Theorem~\ref{Theorem: Gibbs Result} only assumes that the loss function is non-negative, and it holds even for non-i.i.d training samples $S$. 

In Section~\ref{subsec:property}, we discuss some general properties of the expected generalization error that can be derived directly from the properties of symmetrized KL information.  In Section~\ref{sec:mean_example}, we provide a mean estimation example to show that the symmetrized KL information can be computed exactly for squared loss with Gaussian prior. In Section~\ref{sec:other}, we provide some alternative exact characterizations of the expected generalization error using other information measures.

\subsection{General Properties}\label{subsec:property}

By Theorem~\ref{Theorem: Gibbs Result}, some basic properties of the expected generalization error, e.g., non-negativity and concavity, can be proved directly from the properties of symmetrized KL information. 
We also discuss other properties of the symmetrized KL divergence, including data processing inequality,  chain rule, and their implications in learning problems.

\textbf{Non-negativity:} The non-negativity of the expected generalization error, i.e., $\genb(P_{{W}|S}^\gamma,P_S)\ge 0$, follows from the non-negativity of the symmetrized KL information. Note that the non-negativity result could also be proved using \cite[Appendix A.2]{kuzborskij2019distribution} under much more stringent assumptions, including i.i.d samples and a sub-Gaussian loss function.

\textbf{Concavity:} It is shown in \cite{aminian2015capacity} that the symmetrized KL information $I_{\mathrm{SKL}}(X;Y)$ is a concave function of $P_X$ for fixed $P_{Y|X}$, and a convex function of $P_{Y|X}$ for fixed $P_X$. Thus, we have the following corollary.
\begin{corollary}\label{cor:concavity}
 For a fixed $(\gamma,\pi(w),L_\mathrm{e}(w,s))$-Gibbs algorithm $P_{{W}|S}^\gamma$, the expected generalization error $\genb(P_{{W}|S}^\gamma,P_S)$ is a concave function of $P_S$.
\end{corollary}

The concavity of the generalization error for the Gibbs algorithm $P_{{W}|S}^\gamma$ can be immediately used to explain why training a model by mixing multiple datasets from different domains leads to poor generalization. Suppose that the data-generating distribution is domain-dependent, i.e., there exists a random variable $D$, such that $D \leftrightarrow S \leftrightarrow W$ holds. Then, $P_S = \mathbb{E}_{P_D}[P_{S|D}]$ can be viewed as the mixture of the data-generating distribution across all domains. From Corollary ~\ref{cor:concavity} and Jensen's inequality, we have
\begin{equation}
    \genb(P_{{W}|S}^\gamma,P_S) \ge \mathbb{E}_{P_D}\Bigl[ \genb(P_{{W}|S}^\gamma, P_{S|D} )\Bigr],
\end{equation}
which shows that the generalization error of Gibbs algorithm achieved with the mixture distribution $P_S$ is larger than the averaged generalization error for each $P_{S|D}$. 

{\color{black}

\textbf{Lower bound:}
Using Theorem~\ref{Theorem: Gibbs Result} and Pinsker's inequality \cite{palomar2008lautum}, we can also derive the following lower bound on the expected generalization error in terms of total variation distance. As a comparison, an \emph{upper} bound on the generalization error of a learning algorithm in terms of total variation distance is provided in \cite{raginsky2016information}. 
\begin{corollary}\label{Corollary: Lower bound TV}
 For $(\gamma,\pi(w),L_\mathrm{e}(w,s))$-Gibbs algorithm, the following lower bound on the generalization error of the Gibbs algorithm holds:
\begin{equation}
    \genb(P_{{W}|S}^\gamma,P_S) \ge \frac{\TV^2(P_{W,S},P_W\otimes P_S)}{\gamma},
\end{equation}
where 
\begin{align*}
    &\TV(P_{W,S},P_W\otimes P_S)\\
    &\qquad\qquad \triangleq \iint \big|P_{W,S}(w,s)-P_{W}(w)P_S(s)\big|\,\diff w\, \diff s
\end{align*}
denotes total variation distance.
\end{corollary}

Note that the right hand side of the lower bound in Corollary~\ref{Corollary: Lower bound TV} is always bounded in $[0,\frac{4}{\gamma}]$.

\textbf{Upper bound:} We can derive an upper bound on the expected generalization error in terms of symmetrized $\alpha$-R\'enyi divergence. 

\begin{corollary}\label{Corollary: upper bound Renyi}
For $(\gamma,\pi(w),L_\mathrm{e}(w,s))$-Gibbs algorithm, the following upper bound on the generalization error of the Gibbs algorithm holds for $\alpha> 1$: 
\begin{equation}
    \genb(P_{{W}|S}^\gamma,P_S) \le \frac{1}{\gamma}R_{\mathrm{SKL}}^\alpha(P_{W,S},P_W\otimes P_S),
\end{equation}
where
\begin{align*}
    &R_{\mathrm{SKL}}^\alpha(P_{W,S},P_W\otimes P_S) \\&\qquad \triangleq R_\alpha(P_{W,S}\|P_W\otimes P_S)+R_\alpha(P_W\otimes P_S\| P_{W,S})
\end{align*}
\end{corollary}
}

\textbf{Data Processing Inequality:}
As shown in \cite{sason2016f}, symmetrized KL divergence is an $f$-divergence. Thus, the data processing inequality holds, i.e., for Markov chain $S \leftrightarrow W \leftrightarrow W'$,
\begin{equation}
   I_{\mathrm{SKL}}(S;W) \ge I_{\mathrm{SKL}}(S;W').  
\end{equation}
Using the data processing inequality for mutual information, \cite{xu2017information,bu2020information} show that pre/post-processing improves generalization since these techniques give tighter mutual information-based generalization error bounds. 
However, our Theorem~\ref{Theorem: Gibbs Result} only holds for the Gibbs algorithm, which cannot characterize the generalization error for all conditional distributions $P_{W'|S}$ induced by the post-processing $P_{W'|W}$ in the Markov chain. Thus, it is hard to conclude that the pre/post-processing will reduce the exact generalization error for the Gibbs algorithm by directly applying the data processing inequality.

\textbf{Chain Rule:}
As shown in \cite{bu2020information}, using the chain rule of mutual information, i.e., $I(W;S) = \sum_{i=1}^n I(W;Z_i|Z^{i-1})$ and the fact that $I(W;Z_i|Z^{i-1}) \ge I(W;Z_i)$ for i.i.d. samples, the mutual information based generalization bound can be tightened by considering the individual sample mutual information $I(W;Z_i)$.

However, lautum information does not satisfy the same chain
rule \cite{palomar2008lautum} as mutual information in general. Thus, it is hard to characterize the generalization error of the Gibbs algorithm using individual terms $I_{\mathrm{SKL}}(W;Z_i)$.
To see this, we provide an example in Appendix~\ref{App: example for chain-rule} to show that the joint symmetrized KL information $I_{\mathrm{SKL}}(W;S)$ can be either larger or smaller than the sum of individual terms $I_{\mathrm{SKL}}(W;Z_i)$.

\subsection{Example: Mean Estimation}\label{sec:mean_example}
We now consider a simple learning problem, where the symmetrized KL information can be computed exactly to demonstrate the usefulness of Theorem~\ref{Theorem: Gibbs Result}. All details are provided in Appendix~\ref{app:mean}.


Consider the problem of learning the mean $\vmu \in \mathbb{R}^d$ of a random vector $Z$ using $n$ i.i.d training samples $S=\{Z_i\}_{i=1}^n$. We assume that the covariance matrix of $Z$ satisfies  $\Sigma_Z = \sigma^2_Z I_d$ with unknown $\sigma^2_Z$. We adopt the mean-squared loss  $\ell(\vw,\vz) = \|\vz-\vw\|_2^2$, and assume a Gaussian prior for the mean $\pi (\vw) =  \mathcal{N}(\vmu_0,\sigma^2_0 I_d)$. If we set inverse-temperature $\gamma=\frac{n}{2\sigma^2}$, then the $(\frac{n}{2\sigma^2},\mathcal{N}(\vmu_0,\sigma^2_0 I_d),L_\mathrm{e}(\vw,s))$-Gibbs algorithm is given by the following posterior distribution \cite{murphy2007conjugate},
\begin{subequations}
\begin{equation}
    P_{W|S}^{\gamma}(\vw|Z^n) \sim \mathcal{N}\Bigl( \frac{\sigma_1^2 }{\sigma_0^2}\vmu_0 +\frac{\sigma_1^2 }{\sigma^2}\sum_{i=1}^n Z_i,\sigma_1^2 I_d\Bigr),
\end{equation}    
with
\begin{equation}
    \sigma_1^2 = \frac{\sigma_0^2 \sigma^2}{n\sigma_0^2 +\sigma^2}.
\end{equation}
\label{equ:mean_alg}%
\end{subequations}
Since $P_{W|S}^{\gamma}$ is Gaussian, the mutual information and lautum information are given by 
\begin{align}
    &I(S;W) = \frac{n d \sigma_0^2 \sigma_Z^2 }{2(n\sigma_0^2 + \sigma^2)\sigma^2} - D\Bigl(P_W\| \mathcal{N}(\vmu_W,\sigma_1^2 I_d) \Bigr),  \\
    &L(S;W) = \frac{n d \sigma_0^2 \sigma_Z^2 }{2(n\sigma_0^2 + \sigma^2)\sigma^2} + D\Bigl(P_W\| \mathcal{N}(\vmu_W,\sigma_1^2 I_d)),
\end{align}
with
\begin{equation}
\vmu_W = \frac{\sigma_1^2 }{\sigma_0^2}\vmu_0 + \frac{n\sigma_1^2  }{\sigma^2} \vmu.
\end{equation}
For additive Gaussian channel $P_{W|S}$, it is well known that Gaussian input distribution (which also gives a Gaussian output distribution $P_W$) maximizes the  mutual information under a second-order moment constraint.
As we can see from the above expressions, the opposite is true for lautum information. 
In addition, symmetrized KL information $I_{\mathrm{SKL}}(W;S)$
is independent of the distribution of $P_Z$, as long as $\Sigma_Z = \sigma^2_Z I_d$. 


From Theorem~\ref{Theorem: Gibbs Result}, the generalization error of this algorithm can be computed exactly as:
\begin{align}\label{eq:mean_truth}
  \genb(P_{W|S}^{\gamma},P_S) = \frac{I_{\mathrm{SKL}}(W;S)}{\gamma} 
  & = \frac{2 d \sigma_0^2 \sigma_Z^2 }{n\sigma_0^2 + \sigma^2} =\frac{2 d \sigma_0^2 \sigma_Z^2 }{n(\sigma_0^2 +\frac{1}{2\gamma})},
\end{align}
which has the decay rate of $\mathcal{O}\left(1/n\right)$. As a comparison, the individual sample mutual information (ISMI) bound from \cite{bu2020tightening}, which is shown to be tighter than the mutual information-based bound in \cite[Theorem 1]{xu2017information}, gives a sub-optimal bound  with order $\mathcal{O}\left(1/\sqrt{n}\right)$, as $n\to \infty$, (see Appendix \ref{app:ISMI}).
{\color{black}

\subsection{Other Characterizations}\label{sec:other}
In this section, we provide other exact characterizations of the Gibbs algorithm using different information measures. All the proofs can be found in Appendix~\ref{Proof of other characterizations}.

\textbf{Conditional Symmetrized KL Divergence:
}
It is well-known that mutual information has the following variational characterization
\begin{align*}
    I(W;S) &= \inf_{Q_W}D(P_{W|S}\|Q_W|P_S) \\ &=\inf_{Q_W,Q_S}D(P_{W,S}\|Q_W\otimes Q_S),
\end{align*}
which implies that the product-of-marginal distribution minimizes the KL divergence for a given joint distribution.
One may think that the counterpart for lautum information would be $\inf_{Q_W}D(P_S\otimes Q_W\|P_{W,S})$, but it is not true as shown in \cite{palomar2008lautum}. In general, the product-of-marginal distribution does not minimize $D(Q_W\otimes Q_S\|P_{W,S})$, and lautum information satisfies the following variational
characterization
\begin{equation}
    L(W;S) = \inf_{Q_S}D(P_W\otimes P_S\|P_{W|S}\otimes Q_S).
\end{equation}
Thus, the product-of-marginal distribution $P_S \otimes P_W$ does not minimize the symmetrized KL divergence $D_{\mathrm{SKL}}(P_{W,S}\|Q_W\otimes Q_S)$, which implies that there exist different distributions $Q_W$ over the hypothesis space, satisfying the following condition:
\begin{align}\label{eq: rep general}
    I_{\mathrm{SKL}}(W;S)=D(P_{W,S}\|Q_W\otimes P_S) + D(Q_W \otimes P_S \| P_{W,S}).
\end{align}

We characterize the distribution $Q_W$ that satisfies the above equation in the following lemma.  
\begin{lemma}\label{lemma: SKL representation}
For an arbitrary distribution over hypothesis space $Q_W$,  equation \eqref{eq: rep general} holds
if and only if
\begin{equation}\label{Eq: rep main condition1}
    \mathbb{E}_{P_W}[\mathbb{E}_{P_S}[\log(Q_W/P_{W|S})]]=\mathbb{E}_{Q_W}[\mathbb{E}_{P_S}[\log(Q_W/P_{W|S})]].
\end{equation}
\end{lemma}

It can be verified that the condition in Lemma~\ref{lemma: SKL representation},  \eqref{Eq: rep main condition1}, is satisfied, if we consider $(\gamma,\pi(w),L_\mathrm{e}(w,s))$-Gibbs algorithm and $(\gamma,\pi(w),L_\mathrm{p}(w,P_S))$-Gibbs algorithm as $P_{W|S}$ and $Q_W$, respectively.

Thus, inspired by Lemma~\ref{lemma: SKL representation}, we can provide another exact characterization of the expected generalization error of the Gibbs algorithm in terms of symmetrized KL divergence using the distribution $P_{W}^{\gamma,L_\mathrm{p}}$, which denotes the $(\gamma,\pi(w),L_\mathrm{p}(w,P_S))$-Gibbs algorithm. 
\begin{theorem}\label{theorem: Another Representation based on SKL div}
For $(\gamma,\pi(w),L_\mathrm{e}(w,s))$-Gibbs algorithm,
the expected generalization error is given by
\begin{equation}
    \genb(P_{{W}|S}^\gamma,P_{S}) =  \frac{D_{\mathrm{SKL}}(P_{W|S}^{\gamma}\|P_{W}^{\gamma,L_\mathrm{p}}|P_{S})}{\gamma},
\end{equation}
where $D_{\mathrm{SKL}}(P_{W|S}^{\gamma}\|P_{W}^{\gamma,L_\mathrm{p}}|P_S) \triangleq \mathbb{E}_{P_S}[ D_{\mathrm{SKL}}(P_{W|S}^{\gamma}\|P_{W}^{\gamma,L_\mathrm{p}})] $.
\end{theorem}

Using Theorem~\ref{theorem: Another Representation based on SKL div}, we can prove  similar properties, i.e. the non-negativity and concavity of the expected generalization error of the Gibbs algorithm. 

Comparing Theorem~\ref{theorem: Another Representation based on SKL div} with Theorem~\ref{Theorem: Gibbs Result}, we can provide a comparison between the pair of mutual information and $D(P_{W|S}^{\gamma}\|P_{W}^{\gamma,L_\mathrm{p}}|P_{S})$, and the pair of lautum information and $D(P_{W}^{\gamma,L_\mathrm{p}}\| P_{W|S}^{\gamma}|P_{S})$.
\begin{proposition}\label{Prop: compare two bounds}
For $(\gamma,\pi(w),L_\mathrm{e}(w,s))$-Gibbs algorithm we have
\begin{equation}
I(W;S)\leq D(P_{W|S}^{\gamma}\|P_{W}^{\gamma,L_\mathrm{p}}|P_{S}) \text{  and  } L(W;S)\ge D(P_{W}^{\gamma,L_\mathrm{p}}\| P_{W|S}^{\gamma}|P_{S}).
\end{equation}
\end{proposition}

\textbf{Conditional Symmetrized KL Information:}

\color{black}

In this subsection, we consider the supersample method proposed in \cite{steinke2020reasoning} to obtain an exact characterization of the generalization error for the Gibbs algorithm using conditional symmetrized KL information.

Let $\widetilde{S} \in \mathcal{Z}^{n \times 2}$ be a collection of $2n$ samples generated from the data-generating distribution $P_S$, grouped in $n$ pairs, i.e., $\widetilde{S} =\{(\widetilde{Z}_{i,0},\widetilde{Z}_{i,1})\}_{i=1}^n$. Let $U \in \{0,1\}^n$ be $n$ i.i.d uniform Bernoulli random variables, which specify which samples to select from each pair to form the training set, i.e.,  $\widetilde{S}_U = \{ \widetilde{Z} _{i,U_i}\}_{i=1}^n$.

If we consider the following Gibbs algorithm using this random selection process,
\begin{equation}
    P_{{W}|\widetilde{S}, U}^\gamma (w|\widetilde{s},u) = \frac{\pi({w})\, e^{-\gamma L_\mathrm{e}(w,\widetilde{s}_u)}}{V_{L_\mathrm{e}}(\widetilde{s}_u,\gamma)},\quad \gamma>0,
\end{equation}
we have the following result that characterizes the expected generalization error of the $(\gamma,\pi(w),L_\mathrm{e}(w,\widetilde{s}_u))$-Gibbs algorithm in terms of the conditional symmetrized KL information.

\begin{theorem}\label{thm:CMI}
For $(\gamma,\pi(w),L_\mathrm{e}(w,\widetilde{s}_u))$-Gibbs algorithm,
the expected generalization error is given by
\begin{equation}
    \genb(P_{{W}|\widetilde{S}, U}^\gamma,P_{S}) =  \frac{2 I_{\mathrm{SKL}}(W;U|\widetilde{S})}{\gamma}.
\end{equation}
\end{theorem}

\textbf{Replace-one Symmetrized KL Divergence:}
Inspired by the notion of on-average KL-privacy \cite{wang2016average} and \cite[Theorem~1]{kuzborskij2019distribution}, we provide the following characterization of expected generalization error in terms of symmetrized KL divergence between the Gibbs algorithm and one-replace data sample Gibbs algorithm.
\begin{theorem}\label{theorem: Another Representation based on SKL div and stability}
For $(\gamma,\pi(w),L_\mathrm{e}(w,s))$-Gibbs algorithm,
the expected generalization error is given by
\begin{equation}
    \genb(P_{{W}|S}^\gamma,P_{S}) =  \frac{\sum_{i=1}^n D_{\mathrm{SKL}}(P_{W|S}^{\gamma}\|P_{W|S^{(i)}}^{\gamma}|P_{S,Z})}{2\gamma},
\end{equation}
where $S^{(i)}=\{Z_1,\cdots,Z_{i-1},Z,Z_{i+1},\cdots,Z_n \}$ is a replace-one training dataset, i.e., $Z_i$ is replaced by an independent copy $Z$, and $P_{W|S^{(i)}}^{\gamma}$ is the $(\gamma,\pi(w),L_\mathrm{e}(w,s^{(i)}))$-Gibbs algorithm.
\end{theorem}

}

\section{Tighter Generalization Error Upper Bounds}\label{sec:upper_bound}

In this section, we show that by combining the exact characterizations in the previous section, Theorem~\ref{Theorem: Gibbs Result}, Theorem~\ref{theorem: Another Representation based on SKL div}, Theorem~\ref{thm:CMI} and Theorem~\ref{theorem: Another Representation based on SKL div and stability} with existing information-theoretic and PAC-Bayesian approaches, we can provide tighter generalization error upper bounds for the Gibbs algorithm. These bounds quantify how the generalization error of the Gibbs algorithm depends on the number of samples $n$, and are useful when directly evaluating the symmetrized KL information or divergence is difficult.

\subsection{Expected Generalization Error Parametric Upper Bound}

We first provide parametric upper bounds on the expected generalization error for the Gibbs algorithm using previous exact characterizations of generalization error. All the proofs can be found in Appendix~\ref{app: proof Theorem sub Gaussian extension}.

\subsubsection{Parametric Upper Bound via Symmetrized KL Information (Theorem~\ref{Theorem: Gibbs Result})}
The following parametric upper bound on the expected generalization error for the Gibbs algorithm can be obtained by combining our Theorem~\ref{Theorem: Gibbs Result} with the information-theoretic bound proposed in \cite{bu2020tightening} under i.i.d and sub-Gaussian assumptions. A general upper bound on the expected generalization error under bounded CGF assumption is provided in Appendix~\ref{app: proof Theorem sub Gaussian extension}.

\begin{theorem}\label{Theorem: Sub Gaussian extension}
Suppose that the training samples $S=\{Z_i\}_{i=1}^n$ are i.i.d generated from the distribution $P_Z$, and the non-negative loss function $\ell(w,Z)$ is $\sigma$-sub-Gaussian on the left-tail \footnote{A random variable $X$ is $\sigma$-sub-Gaussian on the left-tail if $\log \mathbb{E}[e^{\lambda(X-\mathbb{E}X)}] \le \sigma^2\lambda^2/2$, $\forall \lambda \leq 0 $.}
under distribution $P_Z$ for all $w\in \mathcal{W}$. If we further assume $C_I\le L(W;S)/I(W;S)$ for some $C_I \ge 0$, then for the $(\gamma,\pi(w),L_\mathrm{e}(w,s))$-Gibbs algorithm, we have
\begin{align}
    0\leq\genb(P_{{W}|S}^{\gamma},P_S) \leq \frac{2\sigma^2\gamma}{(1+C_I)n}.
\end{align}
\end{theorem}
Theorem~\ref{Theorem: Sub Gaussian extension} establishes the convergence rate $\mathcal{O}(\gamma/n)$ for the generalization error of Gibbs algorithm with i.i.d training samples, and suggests that a smaller inverse temperature $\gamma$ leads to a smaller generalization error. Note that all the $\sigma$-sub-Gaussian loss functions are also $\sigma$-sub-Gaussian on the left-tail under the same distribution (the mean-squared loss function in Section~\ref{sec:mean_example} is sub-Gaussian on the left-tail under $P_Z$, but not sub-Gaussian). Therefore, our result also applies to any bounded loss function $\ell\colon\mathcal{W}\times \mathcal{Z}\rightarrow [a,b]$, since bounded functions are $(b-a)/2$-sub-Gaussian.
 

\begin{remark}
[Previous Results] Using the fact that Gibbs algorithm is differentially private \cite{mcsherry2007mechanism} for bounded loss functions $\ell \in [0,1]$, directly applying \cite[Theorem~1]{xu2017information} gives a sub-optimal bound  $|\genb(P_{{W}|S}^\gamma,P_S) |\leq \sqrt{\frac{\gamma}{n}}$. By further exploring the bounded loss assumption using Hoeffding’s lemma, a tighter upper bound  $|\genb(P_{{W}|S}^\gamma,P_S) |\leq \frac{\gamma}{2n}$ is obtained in
 \cite{raginsky2016information}, which has the similar decay rate order of $\mathcal{O}\left(\gamma/n\right)$.
In \cite[Theorem~1]{kuzborskij2019distribution}, the upper bound $\genb(P_{{W}|S}^\gamma,P_S) \leq \frac{4 \sigma^2 \gamma}{n}$ is derived with a different assumption, i.e., $\ell(W,z)$ is $\sigma$-sub-Gaussian under Gibbs algorithm $P_{{W}|S}^{\gamma}$. In Theorem~\ref{Theorem: Sub Gaussian extension}, we assume the loss function is $\sigma$-sub-Gaussian on left-tail under data-generating distribution $P_Z$ for all $w\in \mathcal{W}$, which is more general as we discussed above. Our upper bound is also improved by a factor of $\frac{1}{2(1+C_I)}$ compared to the result in \cite{kuzborskij2019distribution}.
\end{remark}



\begin{remark}
[Choice of $C_I$] Since $L(W;S) > 0$ when $I(W;S)>0$, setting $C_I=0$ is always valid in Theorem~\ref{Theorem: Sub Gaussian extension}, which gives $\genb(P_{{W}|S}^\gamma,P_S) \leq \frac{2 \sigma^2 \gamma}{n}$. As shown in \cite[ Theorem~15]{palomar2008lautum}, $L(S;W)\ge I(S;W)$ holds for any Gaussian channel $P_{W|S}$. In addition, it is discussed in \cite[Example~1]{palomar2008lautum}, if either the entropy of training data $S$ or the hypothesis $W$ is small, $I(S;W)$ would be smaller than $L(S;W)$ (as it is not upper-bounded by the entropy), which implies that the lautum information term is not negligible in general.
\end{remark}

We also provide upper bounds for Gibbs algorithm under sub-Exponential and sub-Gamma assumptions, which have the order of $\mathcal{O}(1/n)$ in both cases, in Appendix~\ref{app: More result expected}.

{\color{black}
\subsubsection{Parametric Upper Bound via Symmetrized KL Divergence (Theorem~\ref{theorem: Another Representation based on SKL div})}
We can also combine the following upper bound on the expected generalization error of the Gibbs algorithm in terms of KL divergence with Theorem~\ref{theorem: Another Representation based on SKL div} to provide another parametric upper bound on the Gibbs algorithm under the sub-Gaussian assumption.
\begin{proposition}
\label{Prop: upper on Gibbs based on KL}
Suppose that the training samples $S=\{Z_i\}_{i=1}^n$ are i.i.d generated from the distribution $P_Z$ and the loss function $\ell(w,Z)$ is $\sigma$-sub-Gaussian under distribution $P_Z$ for all $w\in \mathcal{W}$. Then for $(\gamma,\pi(w),L_\mathrm{e}(w,s))$-Gibbs algorithm we have
\begin{align}
    0\leq  \genb(P_{{W}|S}^\gamma,P_S) \leq \sqrt{\frac{2\sigma^2 D(P_{W|S}^{\gamma}\|P_{W}^{\gamma,L_\mathrm{p}}|P_{S}) }{n}}.
\end{align}
\end{proposition}
The following theorem provides another parametric upper bound on expected generalization of the Gibbs algorithm under a different sub-Gaussian assumption.
\begin{theorem}
\label{Theorem: param upper on Gibbs based on KL}
Suppose that the training samples $S=\{Z_i\}_{i=1}^n$ are i.i.d generated from the distribution $P_Z$, and the non-negative loss function $\ell(w,Z)$ is $\sigma$-sub-Gaussian under distribution $P_Z$ for all $w\in \mathcal{W}$. If we further assume $C_K\le D(P_{W}^{\gamma,L_\mathrm{p}}\| P_{W|S}^{\gamma}|P_{S})/D(P_{W|S}^{\gamma}\|P_{W}^{\gamma,L_\mathrm{p}}|P_{S})$ for some $C_K \ge 0$, then for the $(\gamma,\pi(w),L_\mathrm{e}(w,s))$-Gibbs algorithm, we have
\begin{align}
      0\leq  \genb(P_{{W}|S}^\gamma,P_S) \leq \frac{2 \sigma^2 \gamma}{(1+C_K)n}.
\end{align}
\end{theorem}

\begin{remark}[Comparing $C_I$ and $C_K$]
From Proposition~\ref{Prop: compare two bounds}, we can obtain
\begin{align}
    \frac{D(P_{W}^{\gamma,L_\mathrm{p}}\| P_{W|S}^{\gamma}|P_{S})}{D(P_{W|S}^{\gamma}\|P_{W}^{\gamma,L_\mathrm{p}}|P_{S})}\leq\frac{L(W;S)}{I(W;S)}.
\end{align}
Thus, the maximum value of $C_I$ is larger than the maximum value of $C_K$.
\end{remark}

\subsubsection{Parametric Upper Bound via Conditional Symmetrized KL Information (Theorem~\ref{thm:CMI})}

We can combine our Theorem~\ref{thm:CMI} based on conditional symmetrized KL information with generalization error upper bound using conditional mutual information in \cite{steinke2020reasoning}  to provide another parametric upper bound on the Gibbs algorithm under bounded loss condition.

\begin{theorem}\label{Theorem: param via conditional SYI}
Suppose that the training samples $S=\{Z_i\}_{i=1}^n$ are i.i.d generated from the distribution $P_Z$, and the non-negative loss function $\ell(w,z) \in [0,1]$ is bounded. If we further assume $C_C\le L(W;U|\widetilde{S})/I(W;U|\widetilde{S})$ for some $C_C \ge 0$, then for the $(\gamma,\pi(w),L_\mathrm{e}(w,s))$-Gibbs algorithm, we have
\begin{align}
    \genb(P_{{W}|S}^{\gamma},P_S) \leq \frac{\gamma}{(1+C_C)n}.
\end{align}
\end{theorem}

\subsubsection{Parametric Upper Bound via Replace-one Symmetrized KL Divergence
(Theorem~\ref{theorem: Another Representation based on SKL div and stability})}
The following result can be obtained by combining our Theorem~\ref{theorem: Another Representation based on SKL div and stability} and \cite[Theorem~1]{kuzborskij2019distribution}.
\begin{theorem}
\label{Theorem: param upper on Gibbs using stability}
Suppose that the training samples $S=\{Z_i\}_{i=1}^n$ are i.i.d generated from the distribution $P_Z$, and the non-negative loss function $\ell(w,Z)$ is $\tau$-sub-Gaussian under distribution $P_{W|S=s}$ for all $s\in \mathcal{S}$. If we further assume $C_S\le \min_{S^{(i)}}D(P_{W|S^{(i)}}^{\gamma}\| P_{W|S}^{\gamma}|P_{S,Z})/D(P_{W|S}^{\gamma}\|P_{W|S^{(i)}}^{\gamma}|P_{S,Z})$ for some $C_S \ge 0$, then for the $(\gamma,\pi(w),L_\mathrm{e}(w,s))$-Gibbs algorithm, we have
\begin{align}
      0\leq  \genb(P_{{W}|S}^\gamma,P_S) \leq \frac{4 \tau^2 \gamma}{(1+C_S)n}.
\end{align}
\end{theorem}
 
\begin{remark}[$\sigma$-sub-Gaussian Assumption]
The sub-Gaussian assumption in Theorem~\ref{Theorem: param upper on Gibbs using stability} is under the Gibbs algorithm, $P_{W|S=s}^\gamma$ for all $s\in\mathcal{S}$ which is different from the $\sigma$-sub-Gaussian assumption under distribution $P_Z$ for all $w\in\mathcal{W}$ in  Theorem~\ref{Theorem: Sub Gaussian extension} and Theorem~\ref{Theorem: param upper on Gibbs based on KL}.
\end{remark}

We summarized all the exact characterizations of expected generalization error and the tighter expected generalization error upper bounds based on these exact characterizations in Table~\ref{Tab: results}.
\begin{table*}[htb!]
\caption{Exact characterizations of the Gibbs algorithm}
\centering
\begin{tabular}[h!]{|c|c|c|}

 \hline
 Measure & \makecell{Exact Characterization of\\ $\genb(P_{{W}|S}^\gamma,P_{S})$ }& Tighter Expected Upper Bound \\
 \hline\hline
   \makecell{Symmetrized KL Information,\\ (Theorem~\ref{Theorem: Gibbs Result}, Theorem~\ref{Theorem: Sub Gaussian extension})}
 &
  $  \frac{I_{\mathrm{SKL}}(W;S)}{\gamma}$
 &
$\frac{2\sigma^2\gamma}{(1+C_I)n}$
\\
 \hline
 \makecell{Symmetrized KL Divergence,\\ (Theorem~\ref{theorem: Another Representation based on SKL div}, Theorem~\ref{Theorem: param upper on Gibbs based on KL})}
 &  
$  \frac{D_{\mathrm{SKL}}(P_{W|S}^{\gamma}\|P_{W}^{\gamma,L_\mathrm{p}}|P_{S})}{\gamma}$
 &
$\frac{2 \sigma^2 \gamma}{(1+C_K)n}$
\\
 \hline
   \makecell{Conditional Symmetrized KL Information,\\ (Theorem~\ref{thm:CMI}, Theorem~\ref{Theorem: param via conditional SYI})}
 &
 $ \frac{2 I_{\mathrm{SKL}}(W;U|\widetilde{S})}{\gamma}$
&
$\frac{\gamma}{(1+C_C)n}$
\\
 \hline
  \makecell{Replace-one Symmetrized KL Divergence,\\ (Theorem~\ref{theorem: Another Representation based on SKL div and stability}, Theorem~\ref{Theorem: param upper on Gibbs using stability})}
 &
   $\frac{\sum_{i=1}^n D_{\mathrm{SKL}}(P_{W|S}^{\gamma}\|P_{W|S^{(i)}}^{\gamma}|P_{S,Z})}{2\gamma}$
 &
$\frac{4 \tau^2 \gamma}{(1+C_S)n}$
 \\
 \hline
\end{tabular}

\label{Tab: results}
\end{table*}

}

\subsection{PAC-Bayesian Upper Bound}

As discussed in Section~\ref{subsec:related_work}, the prior distribution used in PAC-Bayesian bounds is different from the prior in Gibbs algorithm, since the former  priors can be chosen arbitrarily to tighten the generalization error bound. 
In this section, we provide a tighter PAC-Bayesian bound based on the symmetrized KL divergence as in Theorem~\ref{theorem: Another Representation based on SKL div}, which is inspired by the distribution-dependent PAC-Bayesian bound proposed in \cite{lever2013tighter} using  $(\gamma,\pi(w),L_\mathrm{p}(w,P_S))$-Gibbs distribution as the PAC-Bayesian prior.

As the data-generating distribution $P_S$ is unknown in practice, we consider the $(\gamma,\pi(w),L_\mathrm{p}(w,P_{S^\prime}))$-Gibbs distribution in the following discussion, where  $P_{S^\prime}$ is an arbitrary data-generating distribution.
Since $(\gamma,\pi(w),L_\mathrm{p}(w,P_{S^\prime}))$-Gibbs distribution is independent of the samples $S$ and only depends on the population risk $L_\mathrm{p}(w,P_{S^\prime})$, we can denote it as $P_{W}^{\gamma,L_\mathrm{p}^\prime}$.

By exploiting the connection between the symmetrized KL divergence $D_{\mathrm{SKL}}\Bigl(P_{W|S=s}^{\gamma}\big\|P_{W}^{\gamma,L_\mathrm{p}^\prime}\Bigr) $  and the KL divergence term $D\Bigl(P_{W|S=s}^{\gamma}\big\|P_{W}^{\gamma,L_\mathrm{p}^\prime}\Bigr)$ in the PAC-Bayesian bound from \cite{lever2013tighter}, the following  PAC-Bayesian bound can be obtained under i.i.d and sub-Gaussian assumptions.


\begin{theorem}\label{Theorem: PAC Baysian result}
(proved in Appendix~\ref{app: PAC-Bayesian upper bound}) Suppose that the training samples $S=\{Z_i\}_{i=1}^n$ are i.i.d generated from the distribution $P_Z$, and the non-negative loss function $\ell(w,Z)$ is $\sigma$-sub-Gaussian under data-generating distribution $P_Z$ for all $w\in\mathcal{W}$. If we use the $(\gamma,\pi(w),L_\mathrm{p}(w,P_{Z^\prime}))$-Gibbs distribution as the PAC-Bayesian prior,  where $P_{Z^\prime}$ is an arbitrary chosen (and known) distribution, the following upper bound holds for the generalization error of  $\left(\gamma,\pi(w),L_\mathrm{e}(w,s)\right)$-Gibbs algorithm  with probability at least $1-2\delta$, $0<\delta<1/2$ under distribution $P_S$,
\begin{align*}
    &\left|\mathbb{E}_{P_{W|S=s}^\gamma}[L_\mathrm{p}(W,P_S)-L_\mathrm{e}(W,s)]\right|\\
    &\quad \leq \frac{2\sigma^2\gamma}{(1+C_P(s))n} \\
    &\qquad\quad{}+2\sqrt{\frac{\sigma^2\gamma}{(1+C_P(s))n}}\left(\sqrt[4]{2\sigma^2 D(P_{Z^\prime}\|P_Z)} +\epsilon\right)+ \epsilon^2,
\end{align*}
where 
\begin{equation*}
    \epsilon \triangleq \sqrt[4]{\frac{2\sigma^2\log(1/\delta)}{n}}
    \quad\text{and}\quad C_P(s) \leq \frac{D\Bigl(P_{W}^{\gamma,L_\mathrm{p}^\prime}\big\|P_{W|S=s}^{\gamma}\Bigr)}{D\Bigl(P_{W|S=s}^{\gamma}\big\|P_{W}^{\gamma,L_\mathrm{p}^\prime}\Bigr)},
\end{equation*}
for some $C_P(s)\ge 0$.
\end{theorem}

\begin{remark}[Previous Result] 
We could recover the distribution-dependent bound in \cite[Theorem~6]{lever2013tighter} by setting $P_{Z^\prime} = P_Z$, choosing a bounded loss function in $[0, 1]$ and  $C_P(s)=0$ in our Theorem~\ref{Theorem: PAC Baysian result}. Note that
multiple terms in our upper bound in Theorem~\ref{Theorem: PAC Baysian result} are tightened by a factor of $1/(1+C_P (s))$, and our result applies to $\sigma$-sub-Gaussian loss functions.
\end{remark}
\begin{remark}[Choice of $C_P(s)$]
Since the distribution $P_{Z^\prime}$ can be set arbitrarily, the prior distribution $P_{W}^{\gamma,L_\mathrm{p}^\prime}$ is accessible. Then, we can optimize $C_P(s)={D\Bigl(P_{W}^{\gamma,L_\mathrm{p}^\prime}\big\|P_{W|S=s}^{\gamma}\Bigr)}/{D\Bigl(P_{W|S=s}^{\gamma}\big\|P_{W}^{\gamma,L_\mathrm{p}^\prime}\Bigr)}$ \ to tighten the bound, as it can be computed exactly using the training set.
\end{remark}

\section{Asymptotic Behavior of Generalization Error for Gibbs Algorithm}\label{sec:asymptotic}

In this section, we consider the asymptotic behavior of the generalization error for Gibbs algorithm as the inverse temperature $\gamma \to \infty$\footnote{As discussed in Appendix~\ref{App: EP behaviuor}, with regard to $\gamma$, the expected empirical risk is a decreasing function, therefore it is worthwhile to look into large $\gamma$ behaviour.}. Note that the upper bounds obtained in the previous section, as well as the ones in the literature,  have the order $\mathcal{O}(\frac{\gamma}{n})$, which becomes vacuous in this regime. However, it is known that the Gibbs algorithm will converge to ERM as $\gamma \to \infty$, which has finite generalization error with bounded loss function. To resolve this issue, we provide an exact characterization of the generalization error in this regime using Theorem \ref{Theorem: Gibbs Result}. 



It is shown in \cite{hwang1980laplace,athreya2010gibbs} that the asymptotic behavior of the Gibbs algorithm depends on the number of minimizers for the empirical risk, so we consider the single-well case and multiple-well case separately.

\textbf{Single-well case:}
In this case, there exists a unique $W^*(S)$ that minimizes the empirical risk, i.e.,
\begin{equation}
    W^*(S) = \argmin_{w\in \mathcal{W}} L_\mathrm{e} (w,S).
\end{equation}
It is shown in \cite{hwang1980laplace} that if $H^*(S) \triangleq \nabla^2_w L_\mathrm{e} (w,S)\big|_{w = W^*(S)}$ is not singular, then $P_{W|S}^\gamma \to \mathcal{N}(W^*(S), \frac{1}{\gamma}H^*(S)^{-1})$ 
as $\gamma \to \infty$. Thus, the symmetrized KL information in Theorem~\ref{Theorem: Gibbs Result} can be evaluated using this Gaussian approximation, which gives the following result.

\begin{proposition}\label{prop:single_well} (proved in Appendix~\ref{app: large alpha})
In the single-well case, if the Hessian matrix
$H^*(S)$ is not singular, then the generalization error of the $(\infty,\pi(\vw), L_\mathrm{e}(\vw,s))$-Gibbs algorithm is
\begin{align}
    &\genb(P_{{W}|S}^\infty,P_S) \nn\\
    &\ = \mathbb{E}_{\Delta_{W,S}}\Bigl[\frac{1}{2}W^\top H^*(S) W\Bigr]\\\nn 
    &\qquad{} +\mathbb{E}_{P_{S}}\Bigl[ (W^*(S) - \mathbb{E}[W^*(S)])^\top\\\nn
    &\qquad\qquad\qquad{}\cdot(H^*(S)W^*(S) - \mathbb{E}[H^*(S)W^*(S)]) \Bigr], 
\end{align}
where 
\begin{equation*}
\mathbb{E}_{\Delta_{W,S}}[f(W,S)] \triangleq\mathbb{E}_{P_W\otimes P_S}[f(W,S)] - \mathbb{E}_{P_{W,S}}[f(W,S)].
\end{equation*}
\end{proposition}


Proposition~\ref{prop:single_well} shows that the generalization error of the Gibbs algorithm in the limiting regime $\gamma \rightarrow \infty$ highly depends on the landscape of the empirical risk function.

As an example, we use Proposition~\ref{prop:single_well} to obtain the generalization error of the maximum likelihood estimates (MLE) in the asymptotic regime $n\to \infty$. More specifically,
 suppose that we have $n$ i.i.d. training samples generated from the distribution $P_Z$, and we want to fit the training data with a parametric distribution family $\{f(z_i|\vw)\}_{i=1}^n$, where $\vw\in \mathcal{W}\subset \mathbb{R}^d$ denotes the parameter. Here, the true data-generating distribution may not belong to the parametric family, i.e., $P_Z \ne f(\cdot|\vw)$ for any $\vw \in \mathcal{W}$.  If we use the log-loss $\ell(\vw,z)=-\log f(z|\vw)$ in the Gibbs algorithm, as $\gamma \to \infty$, it converges to the ERM algorithm, which is equivalent to MLE, i.e.,
\begin{equation}
    W^*(S) = \hat{W}_{\mathrm{ML}} \triangleq \argmax_{\vw\in \mathcal{W}} \sum_{i=1}^n \log f(Z_i|\vw).
\end{equation}
As $n\to \infty$, under regularization conditions (details in Appendix~\ref{app:MLE}) which guarantee that $ W^*(S)$ is unique, the asymptotic normality of the MLE \cite{van2000asymptotic} states that the distribution of $\hat{W}_{\mathrm{ML}}$ converges to
\begin{equation}
    \mathcal{N}(\vw^*, \frac{1}{n} J(\vw^*)^{-1} \mathcal{I}(\vw^*)J(\vw^*)^{-1}), 
\end{equation}
with
\begin{align*}
\vw^* &\triangleq \argmin_{\vw\in \mathcal{W}} D(P_Z\|f(\cdot|\vw)),\\
    J(\vw) &\triangleq \E_Z \Bigl[ -\nabla_{\vw}^2 \log f(Z|\vw) \Bigr]
    \end{align*}
and
\begin{align}
    \mathcal{I}(\vw) &\triangleq \E_Z \Bigl[ \nabla_{\vw} \log f(Z|\vw) \nabla_{\vw} \log f(Z|\vw) ^\top\Bigr].
\end{align}
In addition, the Hessian matrix $H^*(S) \to J(\vw^*)$ as $n\to \infty$, which is independent of the training samples $S$. Thus, $\mathbb{E}_{\Delta_{W,S}}[\frac{1}{2}W^\top H^*(S) W]=0$, and Proposition~\ref{prop:single_well} gives 
\begin{equation}\label{eq: AIC}
    \genb(P_{{W}|S}^\infty,P_S) = \frac{\tr( \mathcal{I} (\vw^*) J(\vw^*)^{-1})}{n}.
\end{equation}
When the true model is in the parametric family $P_Z = f(\cdot|\vw^*)$, we have $\mathcal{I}(\vw^*)=J(\vw^*)$ and the above expression reduces to $\genb(P_{{W}|S}^\infty,P_Z) =d/n$, which corresponds to the well-known Akaike information criterion (AIC) \cite{akaike1998information} used in MLE model selection.


In Appendix~\ref{app:bayesian_learning}, we consider a slightly different asymptotic regime,
where the Gibbs algorithm converges to the Bayesian posterior instead of ERM. A similar result as in \eqref{eq: AIC} can be obtained from Bernstein–von–Mises theorem \cite{kleijn2012bernstein} and the asymptotic normality of the MLE.

\textbf{Multiple-well case:}
In this case, there exist $M$ distinct $W_u^*(S)$ such that
\begin{equation}
    W_u^*(S) \in  \argmin_{w\in \mathcal{W}} L_\mathrm{e} (w,S), \quad u \in \{1,\cdots, M\},
\end{equation}
where $M$ is a fixed constant, and all the minimizers $W_u^*(S)$ are isolated,
meaning that a sufficiently small neighborhood of each $W_u^*(S)$ contains a unique minimum.

In this multiple-well case, it is shown in \cite{athreya2010gibbs} that the Gibbs algorithm can be approximated by a Gaussian mixture, as long as $H^*_u(S) \triangleq \nabla^2_w L_\mathrm{e} (w,S)\big|_{w = W_u^*(S)}$ is not singular for all $u\in \{1,\cdots, M\}$. However, there is no closed form for the symmetrized KL information for Gaussian mixtures. Thus,  we provide the following upper bound of the generalization error by evaluating Theorem~\ref{Theorem: Gibbs Result} under the assumption that $\pi(W)$ is a uniform distribution over $\mathcal{W}$.



\begin{proposition}(proved in Appendix~\ref{app: large alpha})\label{prop:multiple_well}
If we assume that $\pi(W)$ is a uniform distribution over $\mathcal{W}$, and the Hessian matrices $H^*_u(S)$ are not singular for all $u\in \{1,\cdots, M\}$, then the generalization error of the $(\infty,\pi(\vw), L_\mathrm{e}(\vw,s))$-Gibbs algorithm in the multiple-well case can be bounded as
\begin{align*}
    &\genb(P_{{W}|S}^{\infty},P_S)\notag\\
    &\quad\le \frac{1}{M}\sum_{u=1}^M \biggl[\mathbb{E}_{\Delta_{W_u,S}}\Bigl[\frac{1}{2}W_u^\top H^*_u(S) W_u\Bigr] \nn\\
    &\qquad\qquad\qquad{}+ \mathbb{E}_{P_{S}}\Bigl[ (W_u^*(S) - \mathbb{E}[W_u^*(S)])^\top H_u \notag\\
    &\qquad\qquad\qquad\qquad\qquad\qquad{}\cdot(W_u^*(S) - \mathbb{E}[W_u^*(S)]) \Bigr] \biggr].
\end{align*}
\end{proposition}

Comparing with Proposition~\ref{prop:single_well}, Proposition~\ref{prop:multiple_well} shows that the global generalization error in the multiple-well case can be upper bounded by the mean of the generalization errors achieved by each local minimizer.

\begin{remark}
In \cite{kuzborskij2019distribution}, a similar Gaussian approximation technique is used to  bound the \emph{excess risk} of Gibbs algorithm in both  single-well and multiple-well cases. However, their result is based on a loose generalization error bound with the order $\mathcal{O}(\frac{\gamma}{n})$. Thus, our method can also be used to obtain a tighter characterization of the excess risk for the Gibbs algorithm.
\end{remark}

\section{Regularized Gibbs Algorithm}\label{sec:regularization}
In this section, we show how regularization will influence the generalization error of the Gibbs algorithm. Our regularizer definition is more general than the standard data-independent regularizer, as it may also depend on the training samples. There are many applications of such data-dependent regularization in the literature---e.g., data-dependent spectral norm regularization proposed in \cite{roth2019adversarial}, $\ell_1$ regularizer over data-dependent hypothesis space  studied in \cite{xiao2010learning} and dropout modeled as data-dependent $\ell_2$ regularization in \cite{wager2013dropout}.

In the following proposition, we consider the Gibbs algorithm with a regularization term  $R:\mathcal{W} \times \mathcal{Z}^n  \to \mathbb{R}_0^+$ and characterize the generalization error of this  $(\gamma,\pi(w),L_\mathrm{e}(w,s) + \lambda R(w,s))$-Gibbs algorithm, which is the solution of the following regularized ERM problem:
\begin{align}\label{Eq: optimization regularization problem}
    P^{\star}_{W|S}=
   \arginf_{P_{W|S}}\biggl( &\mathbb{E}_{P_{W,S}}[L_\mathrm{e}(W,S)+\lambda R(W,S)]\nn\\
   &\quad\qquad{}+\frac{1}{\gamma} D(P_{W|S}\|\pi(W)|P_S)\biggr),
\end{align}
where $\lambda \ge 0$ controls the regularization term. 

\begin{proposition}
\label{Prop: Prop Regularizer}
(proved in Appendix~\ref{app: Regularizer}) For $(\gamma,\pi(w),L_\mathrm{e}(w,s) + \lambda R(w,s))$-Gibbs algorithm, its expected generalization error is given by
\begin{equation}
   \genb(P_{{W}|S}^{\gamma},P_S)=\frac{I_{\mathrm{SKL}}(W;S)}{\gamma}-\lambda\mathbb{E}_{\Delta_{W,S}}[R(W,S)],
\end{equation}
where 
\begin{equation*}
\mathbb{E}_{\Delta_{W,S}}[R(W,S)]=\mathbb{E}_{P_W\otimes P_S}[R(W,S)]-\mathbb{E}_{P_{W,S}}[R(W,S)].
\end{equation*}
\end{proposition}
Proposition~\ref{Prop: Prop Regularizer} holds for non-i.i.d samples and any non-negative loss function, and it shows that in order to improve the generalization ability of the Gibbs algorithm, the data-dependent regularizer needs to 1) minimize the symmetrized KL information $I_{\mathrm{SKL}}(W;S)$ and 2) maximize the $\mathbb{E}_{\Delta_{W,S}}[R(W,S)]$ term which corresponds to a ``generalization error'' defined with the regularization term $R(W,S)$.

\begin{remark}
If the regularizer is independent of the data, i.e., $R(w,s)=R(w)$, we have $\mathbb{E}_{\Delta_{W,S}}[R(W,S)]=0$, and Proposition~\ref{Prop: Prop Regularizer} gives $\genb(P_{{W}|S}^{\gamma},P_S)=\frac{I_{\mathrm{SKL}}(W;S)}{\gamma}$, which implies that the data-independent regularizer needs to improve the generalization ability of learning algorithm by reducing the symmetrized KL information $I_{\mathrm{SKL}}(W;S)$ alone.
\end{remark}

As an example for the data-dependent regularizer, we propose $\ell_2$-regularizer inspired by the regularizer in \cite{shivaswamy2010maximum} for support vector machines. Applying Proposition~\ref{Prop: Prop Regularizer} to this $\ell_2$-regularizer gives the following Corollary.
\begin{corollary}\label{Corollary: l2 regularizer}
(proved in Appendix~\ref{app: Regularizer}) Suppose that we adopt the $\ell_2$-regularizer $R(w,s)=\|w-T(s) \|_2^2$, where $T(\cdot)$ is an arbitrarily deterministic function $T\colon\mathcal{Z}^n \rightarrow \mathcal{W}$. Then, the expected generalization error of $(\gamma,\pi(w),L_\mathrm{e}(w,s) + \lambda R(w,s))$-Gibbs algorithm is
\begin{equation}
  \genb(P_{{W}|S}^{\gamma},P_S)=\frac{I_{\mathrm{SKL}}(W;S)}{\gamma}-\lambda\tr\Bigl(\Cov[W,T(S)]\Bigr),
\end{equation}
where $\Cov[W,T(S)]$ denotes the covariance matrix between $W$ and $T(S)$.
\end{corollary}
The above result suggests that to reduce the generalization error with data-dependent $\ell_2$-regularizer, the function $T(S)$ should be chosen in a way, such that the term $\tr(\Cov[W,T(S)])$ is maximized. One way is to leave a part of the training set and learn the $T(S)$ function. Note that a similar idea has been explored in the development of PAC-Bayesian bound with data-dependent prior \cite{ambroladze2007tighter}.

For general regularization function $R(w,s)$, we can bound the $\mathbb{E}_{\Delta_{W,S}}[R(W,S)]$ term using the mutual information-based generalization error bound in \cite{bu2020tightening,xu2017information}.

\begin{proposition}\label{Proposition: Regularizer general bound}(proved in Appendix~\ref{app: Regularizer})
Suppose that the regularizer function $R(w,s)$  satisfies $ \Lambda_{R(w,s)}(\lambda)\leq \psi(\lambda)$, for $\lambda \in (-b,b)$, $b>0$ under data-generating distribution $P_Z$ for all $w\in \mathcal{W}$. Then the following lower and upper bounds hold for  $(\gamma,\pi(w),L_\mathrm{e}(w,s) + \lambda R(w,s))$-Gibbs algorithm:
\begin{align}\label{eq: reg sub main}
   &\frac{I_{\mathrm{SKL}}(W;S)}{\gamma}-\lambda\psi^{*-1}( I(W;S))\\
   &\qquad\leq \genb(P_{{W}|S}^{\gamma},P_S)\leq \frac{I_{\mathrm{SKL}}(W;S)}{\gamma}+\lambda\psi^{*-1}(I(W;S)). \nn
\end{align}
\end{proposition}

In contrast to the assumption of Theorem~\ref{Theorem: general psi gibs upper bound}, the bounded CGF assumption here is on the regularizer function $R(w,s)$. We could consider different tail behaviors for $R(w,s)$ $\psi(\lambda)$ in Proposition~\ref{Proposition: Regularizer general bound}, including sub-Gaussian, sub-Exponential and sub-Gamma. We provide the bound under the sub-Gaussian assumption in the following corollary for simplicity.
\begin{corollary}
\label{Proposition: Regularizer sub-gaussian}(proved in Appendix~\ref{app: Regularizer})
Suppose that the regularizer function $R(w,s)$ is $\sigma$-sub-Gaussian under the distribution $P_S$ for all $w\in \mathcal{W}$. Then the following bounds holds for  $(\gamma,\pi(w),L_\mathrm{e}(w,s) + \lambda R(w,s))$-Gibbs algorithm:
\begin{align}
   &\frac{I_{\mathrm{SKL}}(W;S)}{\gamma}-\lambda\sqrt{2\sigma^2 I(W;S)}\\
   &\qquad\leq \genb(P_{{W}|S}^{\gamma},P_S)\leq \frac{I_{\mathrm{SKL}}(W;S)}{\gamma}+\lambda\sqrt{2\sigma^2 I(W;S)}.\nn
\end{align}
\end{corollary}

\section{Conclusion}\label{Conc}

{\color{black}In this paper, we provide four different (but equivalent) characterizations of the generalization error for the Gibbs algorithm using symmetrized KL information, symmetrized KL divergence, conditional KL information, and replace-one symmetrized KL divergence, respectively. We demonstrate the power and versatility of our approaches by tightening the expected generalization error using our exact characterizations of generalization error.}

In addition, our information-theoretic method can be applied to provide novel PAC-Bayesian bounds and characterize the behaviors of the Gibbs algorithm with large inverse temperature and the regularized Gibbs algorithm. 

This work motivates further investigation of the Gibbs algorithm in a variety of settings, including extending our results to
characterize the generalization ability of an over-parameterized Gibbs algorithm, which could potentially provide more understanding of the generalization ability for deep learning.

\begin{appendices}
\section{Generalization Error of Gibbs Algorithm}

\subsection{Theorem~\ref{Theorem: Gibbs Result} Details}\label{app: Gibbs Algorithm}
We start with the following two Lemmas:
\begin{lemma}\label{lemma:loss}
We define the following $J_\mathrm{e}(w,S)$ function as a proxy for the empirical risk, i.e., 
\begin{equation}
   J_\mathrm{e}(w,S)\triangleq \frac{\gamma}{n}\sum_{i=1}^n \ell(w,Z_i)+g(w)+h(S), 
\end{equation}
where $\gamma \in \mathbb{R}_0^+$, $g\colon\mathcal{W}\to \mathbb{R}$, $h\colon\mathcal{Z}^n\to \mathbb{R}$, and the function \begin{equation*}
    J_\mathrm{p}(w,P_S)\triangleq \mathbb{E}_{P_S}[J_\mathrm{e}(w,S)]
\end{equation*} 
as a proxy for the population risk.
Then
\begin{equation}
    \mathbb{E}_{P_{W,S}}[J_\mathrm{p}(W,P_S) - J_\mathrm{e}(W,S)] = \gamma \, \genb(P_{W|S},P_S).
\end{equation}
\end{lemma}

\begin{IEEEproof}
\begin{align*}
&\mathbb{E}_{P_{W,S}}[J_\mathrm{p}(W,P_S) - J_\mathrm{e}(W,S)]\nn \\
&=\mathbb{E}_{P_{W,S}}\Bigl[ \mathbb{E}_{P_{S}}[\frac{\gamma}{n}\sum_{i=1}^n \ell(W,Z_i)] -\frac{\gamma}{n}\sum_{i=1}^n \ell(W,Z_i)\Bigr] \nn\\
&\qquad + \mathbb{E}_{P_{W}}\Bigl[g(W)+\mathbb{E}_{P_{S}}[h(S)]\Bigr]- \mathbb{E}_{P_{W,S}}\Bigl[g(W)+ h(S)\Bigr]\\
& = \gamma \, \mathbb{E}_{P_{W,S}}[ L_\mathrm{p}(W,P_S)-L_\mathrm{e}(W,S)] \nn\\
& = \gamma \, \genb(P_{W|S},P_S).\nn 
\end{align*}
\end{IEEEproof}

\begin{lemma}\label{lemma:symmetric_KL}
Consider a learning algorithm $P_{W|S}$, if we set the proxy function $J_\mathrm{e}(w,z^n) = -\log P_{W|S}(w|s)$, 
then
\begin{equation}
    \mathbb{E}_{P_{W,S}}[J_\mathrm{p}(W,P_S) - J_\mathrm{e}(W,S)] = I_{\mathrm{SKL}}(W;S).
\end{equation}
\end{lemma}

\begin{IEEEproof}
\begin{align}
    &I(W;S) + L(W;S) \nn \\
    &= \mathbb{E}_{P_{W,S}}\Bigl[\log \frac{P_{W|S}(W|S)}{P_W(W)}\Bigr] + \mathbb{E}_{P_W \otimes P_S}\Bigl[\log \frac{P_W(W)}{P_{W|S}(W|S)}\Bigr]\nn\\
    &= \mathbb{E}_{P_{W,S}}\Bigl[\log {P_{W|S}(W|S)}\Bigr] - \mathbb{E}_{P_W \otimes P_S}\Bigl[\log {P_{W|S}(W|S)}\Bigr] \\
     &= \mathbb{E}_{P_{W,S}}[-\mathbb{E}_{P_{S}}[\log P_{W|S}(W|S)] +\log P_{W|S}(W|S)]\nn\\
    &=\mathbb{E}_{P_{W,S}}[J_\mathrm{p}(W,P_S) - J_\mathrm{e}(W,S)].\nn 
\end{align}
\end{IEEEproof}

\begin{IEEEproof}[Proof of Theorem~\ref{Theorem: Gibbs Result}]
Considering Lemma~\ref{lemma:loss} and Lemma~\ref{lemma:symmetric_KL}, we just need to verify that $J_\mathrm{e}(w,s) = -\log P_{{W}|S}(w|s)$ can be decomposed into $J_\mathrm{e}(w,s) = \frac{\gamma}{n}\sum_{i=1}^n\ell(w,z_i) +g(w)+h(s)$, for $\gamma>0$. 
Note that
\begin{align}
    J_\mathrm{e}(w,s) 
    &= -\log P_{{W}|S}^\gamma(w|s) \nn\\    
    &=\gamma\, L_\mathrm{e}(w,s) -\log \pi({w})  +\log V_{L_\mathrm{e}}(s,\gamma),
\end{align}
 then we have
\begin{align}
    I_{\mathrm{SKL}}(W;S) &= \mathbb{E}_{P_{W,S}}[J_\mathrm{p}(W,P_S) - J_\mathrm{e}(W,S)]  \\
    &= \gamma \, \genb(P_{{W}|S}^\gamma,P_S). \nn 
\end{align}
\end{IEEEproof}

\begin{IEEEproof}[Proof of Corollary~\ref{Corollary: Lower bound TV}]
This can be proved immediately by combining Theorem~\ref{Theorem: Gibbs Result} with the well-known Pinsker's inequality \cite{palomar2008lautum},
\begin{align}\label{Eq: Pinsker inequality}
    {TV}(P_{W,S},P_W\otimes P_S)\leq \sqrt{2\min(I(W;S),L(W;S))}. 
\end{align}
\end{IEEEproof}

\begin{IEEEproof}[Proof of Corollary~\ref{Corollary: upper bound Renyi}]
First, for $\alpha\geq 1$ we have:
\begin{align}
    &R_{\mathrm{SKL}}^\alpha(P_{W,S},P_W\otimes P_S) \nn\\& = R_\alpha(P_{W,S}\|P_W\otimes P_S)+R_\alpha(P_W\otimes P_S\| P_{W,S})\nn\\
    &\geq D(P_{W,S}\|P_W\otimes P_S) + D(P_W\otimes P_S\| P_{W,S})
\end{align}
where the last inequality is based on the fact that $\alpha$-R\'enyi divergence is an increasing function with respect to $\alpha$ and $R_1(P_{W,S}\|P_W\otimes P_S)=D(P_{W,S}\|P_W\otimes P_S)$.
\end{IEEEproof}

\subsection{Chain-rule and Symmetrized KL Information}\label{App: example for chain-rule}
In the following example, we show that the chain-rule does not hold for symmetrized KL information.
\begin{example}
Consider the following joint distribution for binary random variables $W,Z_1,Z_2 \in \{0,1\}$,
\begin{equation}
    P_{W,Z_1,Z_2}(w,z_1,z_2)=\begin{cases}
\frac{1}{8}, &(z_1,z_2)=(0,0),\\
\frac{1}{4}-\epsilon, & w=1,\ (z_1,z_2)\ne(0,0),\\
\epsilon, &\text{otherwise.}
\end{cases}
\end{equation}
It can be verified that $Z_1$ and $Z_2$ are mutually independent Bernoulli random variable with $p=\frac{1}{2}$, and the conditional distribution is symmetric in the sense that  $P_{W|Z_1,Z_2}(w|0,1) = P_{W|Z_1,Z_2}(w|1,0)$. 

\textbf{Case I:} When $\epsilon=0.0001$, we have
\begin{align*}
I(W;Z_1)&=I(W;Z_2)= 0.0943,\\
I(W;Z_1,Z_2) &= 0.2014, 
\end{align*}
which satisfies the bound 
\begin{equation*}
    I(W;Z_1,Z_2) \ge I(W;Z_1) +I(W;Z_2)
\end{equation*} 
However, we also have
\begin{align*}
    L(W;Z_1)&=L(W;Z_2)= 0.3257,\\L(W;Z_1,Z_2)&=0.5315,
\end{align*}
which satisfies 
\begin{equation*}
L(W;Z_1)+L(W;Z_2)>L(W;Z_1,Z_2), 
\end{equation*}
Hence,
\begin{align*}
I_{\mathrm{SKL}}(W;Z_1)&=I_{\mathrm{SKL}}(W;Z_2)=0.4200,\\ I_{\mathrm{SKL}}(W;Z_1,Z_2)&=  0.7329, 
\end{align*}
and, thus,
\begin{equation*}
I_{\mathrm{SKL}}(W;Z_1)+I_{\mathrm{SKL}}(W;Z_2)>I_{\mathrm{SKL}}(W;Z_1,Z_2).
\end{equation*}

\textbf{Case II:} When $\epsilon=0.01$, it can be verified we have
\begin{align*}
I_{\mathrm{SKL}}(W;Z_1)&=I_{\mathrm{SKL}}(W;Z_2)=0.1255,\\ I_{\mathrm{SKL}}(W;Z_1,Z_2)&=  0.2741,
\end{align*}
and, hence,
\begin{equation*}
I_{\mathrm{SKL}}(W;Z_1)+I_{\mathrm{SKL}}(W;Z_2)<I_{\mathrm{SKL}}(W;Z_1,Z_2).
\end{equation*}
Thus, individual sample  symmetrized KL information cannot be used to characterize the behavior of $I_{\mathrm{SKL}}(W;S)$ in general.
\end{example}

\subsection{Example Details: Mean Estimation }\label{app:mean}
\subsubsection{Generalization Error}
We first evaluate the generalization error of the learning algorithm in \eqref{equ:mean_alg} directly. Note that the output $W$  can be written as
\begin{equation}
    W = \frac{\sigma_1^2 }{\sigma_0^2}\vmu_0 +\frac{\sigma_1^2 }{\sigma^2}\sum_{i=1}^n Z_i +N,\quad\text{with}\quad \sigma_1^2 = \frac{\sigma_0^2 \sigma^2}{n\sigma_0^2 +\sigma^2}
\end{equation}
where $N\sim \mathcal{N}(0,\sigma^2_1 I_d)$ is independent from the training samples $S=\{Z_i\}_{i=1}^n$. Thus,
\begin{align}
   &\genb(P_{W|S},P_S)\nn \\
   &\quad =\mathbb{E}_{P_{W,S}}[ L_\mathrm{p}(W,\mu)-L_\mathrm{e}(W,S)] \nn \\
   &\quad = \mathbb{E}_{P_{W,S}}\biggl[ \mathbb{E}_{P_{\widetilde{Z}}}\Bigl[\|W-\widetilde{Z}\|_2^2\Bigr] - \frac{1}{n}\sum_{i=1}^n\|W-Z_i\|_2^2 \biggr]\nn\\
   &\quad \overset{(a)}{=} \mathbb{E}_{P_{W,Z_i}\otimes P_{\widetilde{Z}}}\Bigl[ (2W-\widetilde{Z}-Z_i)^\top(Z_i - \widetilde{Z}) \Bigr] \nn\\
   &\quad = \mathbb{E}\biggl[ 2\biggl((\frac{\sigma_1^2 }{\sigma_0^2}\vmu_0 +\frac{\sigma_1^2 }{\sigma^2}\sum_{i=1}^n Z_i +N)^\top(Z_i - \widetilde{Z})\biggr) \nn\\
   &\qquad\qquad\qquad\qquad\qquad\qquad{} -(Z_i + \widetilde{Z})^\top(Z_i - \widetilde{Z}) \biggr]\nn\\
   &\quad \overset{(b)}{=} \frac{2\sigma_1^2}{\sigma^2}\mathbb{E}\bigl[  Z_i ^\top(Z_i - \widetilde{Z}) \bigr]\nn\\
   &\quad = \frac{2d\sigma_1^2\sigma_Z^2}{\sigma^2}=\frac{2 d \sigma_0^2 \sigma_Z^2}{n\sigma_0^2 + \sigma^2},
\end{align}
where $\widetilde{Z}\sim \mathcal{N}(\vmu, \sigma_Z^2 I_d)$ denotes an independent copy of the training sample, $(a)$ follows due to the fact that $Z^n$ are i.i.d, and $(b)$ follows from the fact that $Z_i - \widetilde{Z}$ has zero mean, and it is only correlated with $Z_i$.

\subsubsection{Symmetrized KL Divergence}
The following lemma from \cite{palomar2008lautum} characterizes the mutual and lautum information for the Gaussian channel.

\begin{lemma}{\cite[Theorem 14]{palomar2008lautum}}\label{lemma:Gaussian}
Consider the following model 
\begin{equation}
    \mY = \mA \mX+\mN_{\mathrm{G}},
\end{equation}
where $\mX \in \mathbb{R}^{d_X}$ denotes the input
random vector with zero mean (not necessarily
Gaussian), $\mA \in \mathbb{R}^{d_Y \times d_X}$ denotes the linear transformation undergone by the input, $\mY\in \mathbb{R}^{d_Y}$ is the 
output vector, and $\mN_{\mathrm{G}}\in \mathbb{R}^{d_Y}$ is a 
Gaussian noise vector independent of $\mX$. The input and the
noise covariance matrices are given by
$\mSigma$ and $\mSigma_{N_{\mathrm{G}}}$.
Then, we have
\begin{align}
    I(\mX;\mY) &= \frac{1}{2}\tr\Bigl(\mSigma_{N_{\mathrm{G}}}^{-1} \mA \mSigma \mA^\top \Bigr) - D\Bigl(P_\mY\|P_{N_{\mathrm{G}}} \Bigr),  \\
    L(\mX;\mY) &= \frac{1}{2}\tr\Bigl(\mSigma_{N_{\mathrm{G}}}^{-1} \mA \mSigma \mA^\top \Bigr) + D\Bigl(P_\mY\|P_{N_{\mathrm{G}}}\Bigl).
\end{align}

\end{lemma}

In our example, the output $W$ can be written as
\begin{align}
    W &= \frac{\sigma_1^2 }{\sigma_0^2}\vmu_0 +\frac{\sigma_1^2 }{\sigma^2}\sum_{i=1}^n Z_i +N \nn\\
    &= \frac{\sigma_1^2 }{\sigma^2}\sum_{i=1}^n (Z_i-\vmu) + \frac{\sigma_1^2 }{\sigma_0^2}\vmu_0+\frac{n \sigma_1^2  }{\sigma^2}\vmu + N,
\end{align}
where $N\sim \mathcal{N}(0,\sigma^2_1 I_d)$. Setting 
\begin{equation*}
P_{N_{\mathrm{G}}} \sim \mathcal{N}(\frac{\sigma_1^2 }{\sigma_0^2}\vmu_0+\frac{n \sigma_1^2  }{\sigma^2}\vmu, \sigma^2_1 I_d)
\end{equation*}
and $\mSigma = \sigma_Z^2 I_{nd}$ in Lemma~\ref{lemma:Gaussian} gives
\begin{align}
    \tr\Bigl(\mSigma_{N_{\mathrm{G}}}^{-1} \mA \mSigma \mA^\top \Bigr) = \tr\Bigl(\frac{\sigma_Z^2}{\sigma_1^2} \mA  \mA^\top \Bigr),
\end{align}
and noticing that $\mA  \mA^\top = \frac{n \sigma_1^4}{ \sigma^4}I_{d}$  completes the proof.

\subsection{ISMI Bound}\label{app:ISMI}
In this subsection, we  evaluate the following individual sample mutual information (ISMI) bound from \cite[Theorem 2]{bu2020tightening} for the example discussed in Section \ref{sec:mean_example} with i.i.d. samples generated from Gaussian distribution $P_Z \sim \mathcal{N}(\vmu, \sigma_Z^2 I_d)$.

\begin{lemma}{\cite[Theorem 2]{bu2020tightening}}\label{lemma:ISMI}
\quad Suppose $\ell(\widetilde W,\widetilde Z)$ satisfies $\Lambda_{\ell(\widetilde W,\widetilde Z)}(\lambda) \le \psi_{+}(\lambda)$ for $\lambda \in [0,b_+)$, and $\Lambda_{\ell(\widetilde W,\widetilde Z)}(\lambda) \le \psi_{-}(-\lambda)$ for $\lambda\in (b_-,0]$ under $P_{\widetilde Z,\widetilde W} = P_Z\otimes P_{W}$, where $0 <b_+\le \infty$ and $-\infty \leq b_- <0 $. Then,
\begin{align}
    \mathrm{gen}(P_{W|S},P_S) \le \frac{1}{n} \sum_{i=1}^n  \psi^{*-1}_{-}\Bigl(I(W;Z_i)\Bigr),\\
  -\mathrm{gen}(P_{W|S},P_S) \le \frac{1}{n} \sum_{i=1}^n  \psi^{*-1}_{+}\Bigl(I(W;Z_i)\Bigr).
\end{align}
\end{lemma}

We need to compute the mutual information between each individual sample and the output hypothesis $I(W;Z_i)$, and the CGF of $\ell(\widetilde W,\widetilde{Z})$, where $\widetilde{W}$, $\widetilde{Z}$ are independent copies of $W$ and $Z$ with the same marginal distribution, respectively.

Since $W$ and $Z_i$ are Gaussian, $I(W;Z_i)$ can be computed exactly using covariance matrix:
\begin{equation}
 {\rm{Cov}}[Z_i, W] = \left(
    \begin{array}{cc}
     \sigma_Z^2I_d & \frac{\sigma_1^2}{\sigma^2} \sigma_Z^2 I_d  \\
     \frac{\sigma_1^2}{\sigma^2} \sigma_Z^2 I_d &  \Bigl(\frac{n \sigma_1^4}{\sigma^4}\sigma_Z^2 +\sigma_1^2\Bigr)I_d\\
    \end{array}
 \right),
\end{equation}
then, we have
\begin{align}\label{eq:mean_ISMI}
  I(W;Z_i) 
  & = \frac{d}{2} \log \frac{\frac{n \sigma_1^4}{\sigma^4}\sigma_Z^2 +\sigma_1^2}{\frac{(n-1) \sigma_1^4}{\sigma^4}\sigma_Z^2 +\sigma_1^2} \nn \\
  &= \frac{d}{2} \log\Bigl(1+ \frac{\sigma_1^2\sigma_Z^2}{(n-1) \sigma_1^2\sigma_Z^2 +\sigma^4}\Bigr)\nn \\
  & = \frac{d}{2} \log\Bigl(1+ \frac{\sigma_0^2\sigma_Z^2}{(n-1) \sigma_0^2\sigma_Z^2 +n \sigma_0^2 \sigma^2+\sigma^4}\Bigr),
\end{align}
for $i=1,\cdots,n$, $n\ge 2$. 
In addition, since
\begin{equation}
    W\sim \mathcal{N}\Bigl(\frac{\sigma_1^2 }{\sigma_0^2}\vmu_0 +\frac{n\sigma_1^2 }{\sigma^2}\vmu, \Bigl(\frac{n \sigma_1^4}{\sigma^4}\sigma_Z^2 +\sigma_1^2\Bigr)I_d\Bigr),
\end{equation}
it can be shown that $\ell(\widetilde W,\widetilde{Z})=\|\widetilde{Z}-\widetilde{W}\|^2$ is a scaled non-central chi-square distribution with $d$ degrees of freedom, where the scaling factor $\sigma_{\ell}^2 \triangleq (\frac{n\sigma_1^4}{\sigma^4}+1)\sigma_Z^2+\sigma_1^2$ and its non-centrality parameter $\eta \triangleq \frac{\sigma^2}{n \sigma_0^2+\sigma^2}\|\vmu_0-\vmu\|_2^2$. 

Note that the expectation of chi-square distribution with non-centrality parameter $\eta$ and $d$ degrees of freedom is $d + \eta$ and its moment generating function is $ \exp(\frac{\eta \lambda}{1-2\lambda})(1-2\lambda)^{-d/2}$. Therefore, the CGF of $\ell(\widetilde W,\widetilde Z)$ is given by
\begin{equation}
  \Lambda_{\ell(\widetilde W,\widetilde Z)}(\lambda) = - (d \sigma_\ell^2 +\eta) \lambda +\frac{\eta\lambda}{1-2\sigma_\ell^2 \lambda}- \frac{d}{2} \log(1-2\sigma_\ell^2\lambda),
\end{equation}
for $\lambda \in (-\infty, \frac{1}{2\sigma_\ell^2})$.
Since $\mathrm{gen}(P_{W|S},P_Z)\ge 0$, we only need to consider the case $\lambda <0$. It can be shown that:
\begin{align}
\Lambda_{\ell(\widetilde W,\widetilde Z)}(\lambda) &= - d \sigma_\ell^2 \lambda - \frac{d}{2} \log(1-2\sigma_\ell^2\lambda) +\frac{2\sigma_\ell^2\eta\lambda^2}{1-2\sigma_\ell^2\lambda} \nn \\
& = \frac{d}{2} (-u-\log(1-u)) +\frac{2\sigma_\ell^2\eta\lambda^2}{1-2\sigma_\ell^2\lambda},
\end{align}
where $u\triangleq2\sigma_\ell^2\lambda$. Further note  that
\begin{align}
-u-\log(1-u) \le \frac{u^2}{2},\ u&<0,\\
\frac{2\sigma_\ell^2\eta\lambda^2}{1-2\sigma_\ell^2\lambda} \le 2\sigma_\ell^2\eta\lambda^2,\ \lambda&<0.
\end{align}
We have the following upper bound on the CGF of $\ell(\widetilde{W},\widetilde{Z})$:
\begin{equation}
\Lambda_{\ell(\widetilde W,\widetilde Z)}(\lambda) \le (d\sigma_\ell^4+2\sigma_\ell^2\eta)\lambda^2,\quad \lambda<0,
\end{equation}
which means that $\ell(\widetilde W,\widetilde Z)$ is $\sqrt{d\sigma_\ell^4+2\sigma_\ell^2\eta}$-sub-Gaussian for $\lambda<0$.
Combining the results in \eqref{eq:mean_ISMI}, Lemma~\ref{lemma:ISMI} gives the following bound
\begin{align}\label{equ:ISMI_bound}
   &\genb(P_{W|S},P_S)\\
   &\ \le \sqrt{\frac{d^2 \sigma_{\ell}^4 +2d \sigma_{\ell}^2\eta}{2} \log(1+ \frac{\sigma_0^2\sigma_Z^2}{(n-1) \sigma_0^2\sigma_Z^2 +n \sigma_0^2 \sigma^2+\sigma^4})}.
\end{align}
If $\sigma^2=\frac{n}{2\gamma}$ is a constant, i.e., $\gamma = \mathcal{O}(n)$, then as $n \to \infty$, $\sigma_{1}^2 = \mathcal{O}\Bigl(\frac{1}{n}\Bigr)$ and $\sigma_{\ell}^2 = \mathcal{O}(1)$, and the above bound is $\mathcal{O}\left(\frac{1}{\sqrt{n}}\right)$.

\subsection{Other Characterizations}\label{Proof of other characterizations}

\begin{IEEEproof}[Proof of Lemma~\ref{lemma: SKL representation}]
Consider $Q_W$ as an arbitrary distribution on hypothesis space, then the variational representations of mutual information and lautum information are given by
\begin{align}
    &I(W;S)=D(P_{W,S}\|Q_W\otimes P_S)-D(P_W\|Q_W)\\
    &L(W;S)=\mathbb{E}_{P_S\otimes P_W}[\log(Q_W/P_{W|S})]+D(P_W\|Q_W).
\end{align}
Now for $I_{\mathrm{SKL}}(W;S)$ we have
\begin{align}\label{eq: rep 1}
    &I_{\mathrm{SKL}}(W;S)\nn\\
    &\qquad=I(W;S)+L(W;S)\\
    &\qquad= D(P_{W,S}\|Q_W\otimes P_S)+ \mathbb{E}_{P_S\otimes P_W}[\log(Q_W/P_{W|S})],\nn
\end{align}
which is valid for all $Q_W$. We compare this representation with the following
\begin{align}
    D(P_{W,S}\|Q_W\otimes P_S) + D(Q_W \otimes P_S \| P_{W,S}).
\end{align}
The difference between these two expressions is 
\begin{align}
    &I_{\mathrm{SKL}}(W;S)-\left(D(P_{W,S}\|Q_W\otimes P_S) + D(Q_W \otimes P_S \| P_{W,S})\right)\nn\\
    &=\mathbb{E}_{P_S\otimes P_W}[\log(Q_W/P_{W|S})]-D(Q_W \otimes P_S \| P_{W,S})\nn\\
    &=\mathbb{E}_{P_S\otimes P_W}[\log(Q_W/P_{W|S})]-\mathbb{E}_{P_S\otimes Q_W}[\log(Q_W/P_{W|S})]\nn\\
    &=\mathbb{E}_{\Delta(P_W-Q_W)}[\mathbb{E}_{P_S}[\log(Q_W/P_{W|S})]].
\end{align}
If $Q_W$ satisfies the following condition
\begin{equation}
    \mathbb{E}_{\Delta(P_W-Q_W)}[\mathbb{E}_{P_S}[\log(Q_W/P_{W|S})]]=0,
\end{equation}
then we have
\begin{align}
    I_{\mathrm{SKL}}(W;S)=D(P_{W,S}\|Q_W\otimes P_S) + D(Q_W \otimes P_S \| P_{W,S}).
\end{align}
\end{IEEEproof}

 The following lemma  provides an operational interpretation of the symmetrized KL divergence between the Gibbs posterior $P_{W|S}^{\gamma}$ and the prior distribution $P_{W}^{\gamma,L_\mathrm{p}^\prime}$. 

\begin{lemma}\label{Lemma: exact SKL two Gibbs} 
Let us denote the $(\gamma,\pi(w),L_\mathrm{e}(w,s))$-Gibbs algorithm as $P_{W|S}^{\gamma}$ and the $(\gamma,\pi(w),L_\mathrm{p}(w,P_{S^\prime}))$-Gibbs algorithm as $P_{W}^{\gamma,L_\mathrm{p}^\prime}$. Then, the following equality holds for these two Gibbs distributions with the same inverse temperature and prior distribution
\begin{align}\label{equ:SKL_divergence-f}
&\mathbb{E}_{\Delta(P_{W|S=s}^{\gamma},P_{W}^{\gamma,L_\mathrm{p}^\prime})}[L_\mathrm{p}(W,P_{S^\prime})-L_\mathrm{e}(W,s)]\nn\\
&\qquad\qquad\qquad\qquad\qquad =\frac{D_{\mathrm{SKL}}(P_{W|S=s}^{\gamma}\|P_{W}^{\gamma,L_\mathrm{p}^\prime})}{\gamma},
\end{align}
where 
\begin{equation*}
\mathbb{E}_{\Delta(P_{W|S=s}^{\gamma},P_{W}^{\gamma,L_\mathrm{p}^\prime})}[f(W)]=\mathbb{E}_{P_{W|S=s}^{\gamma}}[f(W)]-\mathbb{E}_{P_{W}^{\gamma,L_\mathrm{p}^\prime}}[f(W)].
\end{equation*}
\end{lemma}

\begin{IEEEproof}
\begin{align}
    &D_{\mathrm{SKL}}(P_{W|S=s}^{\gamma}\|P_{W}^{\gamma,L_\mathrm{p}^\prime}) \nn \\
    &= \mathbb{E}_{P_{W|S=s}^{\gamma}} \left[\log\frac{P_{W|S=s}^{\gamma}}{P_{W}^{\gamma,L_\mathrm{p}^\prime}}\right]-\mathbb{E}_{P_{W}^{\gamma,L_\mathrm{p}^\prime}}\left[ \log\frac{P_{W|S=s}^{\gamma}}{P_{W}^{\gamma,L_\mathrm{p}^\prime}}\right]\nn\\
    &\overset{(a)}{=} \mathbb{E}_{\Delta(P_{W|S=s}^{\gamma},P_{W}^{\gamma,L_\mathrm{p}^\prime})} \Bigl[\log(e^{-\gamma (L_\mathrm{e}(W,s)-L_\mathrm{p}(W,P_{S^\prime}))})\Bigr]\nn\\
    &=\gamma\  \mathbb{E}_{\Delta(P_{W|S=s}^{\gamma},P_{W}^{\gamma,L_\mathrm{p}^\prime})}\Bigl[L_\mathrm{p}(W,P_{S^\prime})-L_\mathrm{e}(W,s)\Bigr],
\end{align}
where (a) follows by the fact that partition functions $V_{L_\mathrm{e}}(s,\gamma)$ do not depend on $W$.
\end{IEEEproof}

\begin{IEEEproof}[Proof of Theorem~\ref{theorem: Another Representation based on SKL div}]
In Lemma~\ref{Lemma: exact SKL two Gibbs}, if we consider $P_{S^\prime}=P_S$ and take expectation over $P_S$ in \eqref{equ:SKL_divergence-f} and notice the fact that $\mathbb{E}_{P_{W}^{\gamma,L_\mathrm{p}}\otimes P_{S}}[L_\mathrm{e}(W,S)]= \mathbb{E}_{P_{W}^{\gamma,L_\mathrm{p}}}[L_\mathrm{p}(W,P_{S})]$, we obtain a characterization of the  expected generalization error in terms of the symmetrized KL divergence, i.e., 
\begin{equation}\label{Eq:gen in terms of skl div}
    \genb(P_{W|S}^{\gamma},P_S) =\frac{D_{\mathrm{SKL}}(P_{W|S}^{\gamma}\|P_{W}^{\gamma,L_\mathrm{p}}|P_S)}{\gamma}.
\end{equation}
\end{IEEEproof}

{\color{black}
\begin{IEEEproof}[Proof of Proposition~\ref{Prop: compare two bounds}]

From Theorem~\ref{theorem: Another Representation based on SKL div} and Theorem~\ref{Theorem: Gibbs Result}, we have the following equation for $(\gamma,\pi(w),L_\mathrm{e}(w,s))$-Gibbs algorithm, 
\begin{align}\label{Eq: equality of two reps}
    &I(W;S)+L(W;S)\\\nonumber
    &=D(P_{W|S}^{\gamma}\|P_{W}^{\gamma,L_\mathrm{p}}|P_{S})+ D(P_{W}^{\gamma,L_\mathrm{p}}\| P_{W|S}^{\gamma}|P_{S}).
\end{align}
Note that mutual information has the following variational representation:
\begin{align}\label{Eq: rep mutual}
    I(W;S)=D(P_{W,S}\|Q_W\otimes P_S)-D(P_W\|Q_W).
\end{align}
Let $Q_W=P_{W}^{\gamma,L_\mathrm{p}}$ in \eqref{Eq: rep mutual}, we have
\begin{align}\label{Eq: mutual bound rep}
    I(W;S)\leq D(P_{W|S}^{\gamma}\|P_{W}^{\gamma,L_\mathrm{p}}|P_{S}).
\end{align}
Using \eqref{Eq: equality of two reps} and \eqref{Eq: mutual bound rep}, the following holds
\begin{align}
     L(W;S)\geq D(P_{W}^{\gamma,L_\mathrm{p}}\| P_{W|S}^{\gamma}|P_{S}).
\end{align}
\end{IEEEproof}

\begin{IEEEproof}[Proof of Theorem~\ref{thm:CMI}]
\begin{align}
    &I_{\mathrm{SKL}}(W;U|\widetilde{S}) \nn \\
    &= \mathbb{E}_{P_{\widetilde{S}}} \Big[ \mathbb{E}_{P_{W,U|\widetilde{S}}}\big[\log \frac{P_{W|\widetilde{S},U}^\gamma}{P_{W|\widetilde{S}}}\big]+\mathbb{E}_{P_{W|\widetilde{S}} \otimes P_{U|\widetilde{S}}  }\big[\log \frac{P_{W|\widetilde{S}}}{P_{W|\widetilde{S},U}^\gamma}\big] \Big] \nn \\
    & = \mathbb{E}_{P_{\widetilde{S}}} \Big[ \mathbb{E}_{P_{W,U|\widetilde{S}}}\big[\log \frac{P_{W|\widetilde{S},U}^\gamma}{P_{W|\widetilde{S}}}\big]-\mathbb{E}_{P_{W|\widetilde{S}} \otimes P_{U}  }\big[\log \frac{P_{W|\widetilde{S},U}^\gamma}{P_{W|\widetilde{S}}}\big] \Big]\nn \\
    & = \mathbb{E}_{P_{\widetilde{S}}} \Big[ \mathbb{E}_{P_{W,U|\widetilde{S}}}\big[\log {P_{W|\widetilde{S},U}^\gamma}\big]-\mathbb{E}_{P_{W|\widetilde{S}} \otimes P_{U}  }\big[\log {P_{W|\widetilde{S},U}^\gamma}\big] \Big]\nn \\
    & = \gamma \mathbb{E}_{P_{\widetilde{S}}} \Big[ \mathbb{E}_{P_{W,U|\widetilde{S}}}\big[ L_\mathrm{e}(W,\widetilde{S}_U) \big]-\mathbb{E}_{P_{W|\widetilde{S}} \otimes P_{U}  }\big[ L_\mathrm{e}(W,\widetilde{S}_U) \big]\Big]\nn\\
    &\overset{(a)}{=} \gamma  \mathbb{E}_{P_{W,U,\widetilde{S}}}\big[ L_\mathrm{e}(W,\widetilde{S}_U) \big] \nn \\ &\qquad -\frac{\gamma}{2}\Big[\mathbb{E}_{P_{W,U,\widetilde{S}}}\big[ L_\mathrm{e}(W,\widetilde{S}_U)\big] + L_\mathrm{p}(W,P_S) \Big]\nn\\
    & = \frac{\gamma}{2} \genb(P_{{W}|S}^\gamma,P_{S}).
\end{align}
where $(a)$ follows from the fact that if we independently draw $n$ samples from $\widetilde{S}$, on average $\frac{n}{2}$ of them will be training samples, and the remain $\frac{n}{2}$ samples are test samples. 
\end{IEEEproof}

\begin{IEEEproof}[Proof of Theorem~\ref{theorem: Another Representation based on SKL div and stability}]
As shown in the proof of \cite[Theorem~1]{kuzborskij2019distribution},
\begin{align}\label{eq: gen stability rep}
    &\genb(P_{W|S}^{\gamma},P_S) \\
    &= \frac{1}{n}\sum_{i=1}^n\left(\mathbb{E}_{P_{S,Z}}\left[\mathbb{E}_{P_{W|S^{(i)}}}[\ell(W,Z_i)]-\mathbb{E}_{P_{W|S}}[\ell(W,Z_i)]\right]\right). \nn
\end{align}
The conditional symmetrized KL divergence, i.e., $D_{\mathrm{SKL}}(P_{W|S}^{\gamma}\|P_{W|S^{(i)}}^{\gamma}|P_{S,Z})$, can be represented as follows
\begin{align}
    &D_{\mathrm{SKL}}(P_{W|S}^{\gamma}\|P_{W|S^{(i)}}^{\gamma}|P_{S,Z})\nn \\
    &=\mathbb{E}_{P_{S,Z}}[D(P_{W|S}^{\gamma}\|P_{W|S^{(i)}}^{\gamma})+D(P_{W|S^{(i)}}^{\gamma}\|P_{W|S}^{\gamma})]
    \\& =\mathbb{E}_{P_{S,Z}}[\frac{\gamma}{n}(\mathbb{E}_{P_{W|S}}[\ell(W,Z)]-\mathbb{E}_{P_{W|S}}[\ell(W,Z_i)])]\nn\\
    &    \quad + \mathbb{E}_{P_{S,Z}}[\frac{\gamma}{n}(\mathbb{E}_{P_{W|S^{(i)}}}[\ell(W,Z_i)]-\mathbb{E}_{P_{W|S^{(i)}}}[\ell(W,Z)])] \nn
\end{align}
Due to the fact that $Z$ and $Z_i$ are exchangeable, we have
\begin{align}
    &\mathbb{E}_{P_{S,Z}}\left[\mathbb{E}_{P_{W|S^{(i)}}}[\ell(W,Z_i)]-\mathbb{E}_{P_{W|S}}[\ell(W,Z_i)]\right]\\\nonumber&
   = \mathbb{E}_{P_{S,Z}}\left[\mathbb{E}_{P_{W|S}}[\ell(W,Z)]-\mathbb{E}_{P_{W|S^{(i)}}}[\ell(W,Z)]\right], 
\end{align}
the final result holds.
\end{IEEEproof}
}

\section{Expected Generalization Error Upper Bound}\label{app: Average Upper Bound Details}

\subsection{Preliminaries}\label{subsec:CGF}
To present the existing information-theoretic generalization error bounds, we start with the introduction of the cumulant generating function, which characterizes different tail behaviors of random variables. 
\begin{definition}
The cumulant generating function (CGF) of a random variable $X$ is defined as
\begin{equation}
\Lambda_X(\lambda) \triangleq \log \mathbb{E}[e^{\lambda(X-\mathbb{E}X)}].
\end{equation}
\end{definition}
Assuming $\Lambda_X(\lambda)$ exists, it can be verified that $\Lambda_X(0)=\Lambda_X'(0)=0$, and that it is convex.

\begin{definition}
For a convex function $\psi$ defined on the interval $[0,b)$, where $0<b\le \infty$, its Legendre dual $\psi^\star$ is defined as
\begin{equation}
  \psi^\star(x) \triangleq \sup_{\lambda \in [0,b)} \Bigl(\lambda x-\psi(\lambda)\Bigr).
\end{equation}
\end{definition}

The following lemma characterizes a useful property of the Legendre dual and its inverse function.
\begin{lemma}\label{lemma:psi_star}\cite[Lemma 2.4]{boucheron2013concentration} 
Assume that $\psi(0)= \psi'(0) = 0$. Then $\psi^\star(x)$ defined above is a non-negative convex and non-decreasing function on $[0,\infty)$ with $\psi^\star(0) = 0$. Moreover, its inverse function $\psi^{\star -1}(y) = \inf\{x\ge 0\colon\psi^\star(x)\ge y\}$ is concave, and can be written as
\begin{equation}
  \psi^{\star -1}(y) = \inf_{\lambda \in [0,b)} \Bigl( \frac{y+\psi(\lambda)}{\lambda} \Bigr),\quad b> 0.
\end{equation}
\end{lemma}
Here, we consider the distributions with the following tail behaviors:
\begin{itemize}
\item \textbf{Sub-Gaussian:} A random variable $X$ is  $\sigma$-sub-Gaussian, if $\psi(\lambda)= \frac{\sigma^2\lambda^2}{2}$ is an upper bound of $\Lambda_X(\lambda)$, for $\lambda \in \mathbb{R}$. Then by Lemma~\ref{lemma:psi_star}, $$\psi^{\star -1}(y)=\sqrt{2\sigma^2y}.$$
\item \textbf{Sub-Exponential:} A random variable $X$ is  $(\sigma_e^2,b)$-sub-Exponential, if $\psi(\lambda)= \frac{\sigma_e^2\lambda^2}{2}$ is an upper bound of $\Lambda_X(\lambda)$, for $0 \leq |\lambda|\leq \frac{1}{b} $ and $b > 0$. By Lemma~\ref{lemma:psi_star}, we have 
$$
\psi^{\star -1}(y)=
    \begin{cases}                       &\sqrt{2\sigma_e^2y}, \quad y\leq \frac{\sigma_e^2}{2b};\\
    &by+\frac{\sigma_e^2}{2b},\quad \textit{otherwise.}
    \end{cases}
$$
\item \textbf{Sub-Gamma:} A random variable $X$ is  $\Gamma(\tau^2,c_s)$-sub-Gamma \cite{zhang2020concentration}, if $\psi(\lambda)=  \frac{\lambda^2 \tau^2}{2(1-c_s |\lambda|)}$ is an upper bound of $\Lambda_X(\lambda)$, for $0 < |\lambda| < \frac{1}{c_s} $ and $c_s>0$. By Lemma~\ref{lemma:psi_star}, we have $$\psi^{\star -1}(y)=\sqrt{2\tau^2y}+c_s y.$$
\end{itemize}

The sub-Exponential condition is slightly milder compared with the sub-Gaussian condition. All the definitions above can be generalized by considering only the left ($\lambda <0$) or right ($\lambda >0$) tails, e.g., $\sigma$-sub-Gaussian in the left tail. 

\subsection{Proofs}
\label{app: proof Theorem sub Gaussian extension}

We prove a slightly more general form of Theorem~\ref{Theorem: Sub Gaussian extension} as follows:

\begin{theorem}\label{Theorem: general psi gibs upper bound}
Suppose that the training samples $S=\{Z_i\}_{i=1}^n$ are i.i.d generated from the distribution $P_Z$ and the loss function $\ell(w,Z)$ satisfies $ \Lambda_{\ell(w,Z)}(\lambda)\leq \psi(-\lambda)$, for $\lambda \in (-b,0)$, $b>0$ under data-generating distribution $P_Z$ for all $w\in \mathcal{W}$. Let us assume there exists $C_I\in \mathbb{R}_0^+$ such that $\frac{L(W;S)}{I(W;S)}\geq C_I$, and we further assume there exists $0<\kappa<\infty$ such that
\begin{equation}
 \psi^{\star -1}\left(\frac{\kappa}{n}\right) - \frac{(1+C_I)\kappa}{\gamma}=0.
\end{equation} 
Then, the following upper bound holds for the expected generalization error of $(\gamma,\pi(w),L_\mathrm{e}(w,s))$-Gibbs algorithm:
\begin{equation}
    0\leq \genb(P_{{W}|S}^{\gamma},P_S) \leq \frac{(1+C_I)\kappa}{\gamma}.
\end{equation}
\end{theorem}

\begin{IEEEproof}[Proof of Theorem~\ref{Theorem: general psi gibs upper bound}]
It is shown in \cite[Proposition 2]{bu2020tightening} that the following generalization error bound holds,
\begin{equation}
    \genb(P_{{W}|S}^{\gamma},P_S) \leq \psi^{\star -1}\Bigl(\frac{I(W;S)}{n}\Bigr).
\end{equation}
By Theorem~\ref{Theorem: Gibbs Result} and the assumption on $C_I$, we have
\begin{equation}
  \genb(P_{{W}|S}^{\gamma},P_S) = \frac{ I(W;S) +L(W;S)}{\gamma} \ge \frac{(1+C_I) I(W;S)}{\gamma}.
\end{equation}
Therefore, 
\begin{equation}
  \frac{(1+C_I)I(W;S)}{\gamma}  \leq \psi^{\star -1}\Bigl(\frac{I(W;S)}{n}\Bigr).
\end{equation}
Consider the function $F(u)\triangleq \psi^{\star -1}(\frac{u}{n})-\frac{(1+C_I)u}{\gamma}$, which is concave and satisfies $F(0)=0$ by Lemma~\ref{lemma:psi_star}. If there exists $0<\kappa<\infty$, such that $ F(\kappa)=0 $, then $F(I(W;S))\ge 0$ implies that
$$ 0\leq I(W;S)\leq\kappa. $$

Since $\psi^{\star -1}(\cdot)$ is non-decreasing, we have
\begin{equation*}
    \genb(P_{{W}|S}^{\gamma},P_S) \leq \psi^{\star -1}\Bigl(\frac{\kappa}{n}\Bigr) = \frac{(1+C_I)\kappa}{\gamma}. 
\end{equation*}

\end{IEEEproof}

Note that Theorem~\ref{Theorem: general psi gibs upper bound} can be applied to the cases where the loss functions have different tail distributions discussed in Section~\ref{subsec:CGF}. However, the upper bound in \cite[Theorem~1]{kuzborskij2019distribution} is only applicable with sub-Gaussian assumption.

We can specify the different forms of 
$\psi(\lambda)$ function in Theorem~\ref{Theorem: general psi gibs upper bound} to capture different tail behaviors of the loss function. We first consider the $\sigma$-sub-Gaussian assumption.

\begin{IEEEproof}[Proof of Theorem~\ref{Theorem: Sub Gaussian extension}]
If the loss function is $\sigma$-sub-Gaussian on the left-tail we have $\psi^{\star -1}(y)=\sqrt{2\sigma^2 y}$. Using Theorem~\ref{Theorem: general psi gibs upper bound} we have
\begin{equation}
    \sqrt{2\sigma^2\frac{\kappa}{n}} - \frac{(1+C_I)\kappa}{\gamma}=0,
\end{equation} 
and the solution is $\kappa=\frac{2\sigma^2}{n}\frac{\gamma^2 }{(1+C_I)^2}$. Therefore, 
\begin{equation*}
    \genb(P_{{W}|S}^{\gamma},P_S) \leq \frac{(1+C_I)\kappa}{\gamma}=\frac{2\sigma^2 \gamma}{n(1+C_I)}. 
\end{equation*}
\end{IEEEproof}
{\color{black}
\begin{IEEEproof}[Proof of Proposition~\ref{Prop: upper on Gibbs based on KL}]
From Proposition~\ref{Prop: compare two bounds}, we have
\begin{align}
      I(W;S)\leq D(P_{W|S}^{\gamma}\|P_{W}^{\gamma,L_\mathrm{p}}|P_{S}).
\end{align}
Substituting the mutual information with $D(P_{W|S}^{\gamma}\|P_{W}^{\gamma,L_\mathrm{p}}|P_{S})$ in \cite[Theorem~1]{xu2017information}, the final result holds.
\end{IEEEproof}

\begin{IEEEproof}[Proof of Theorem~\ref{Theorem: param upper on Gibbs based on KL}]
Combining Theorem~\ref{theorem: Another Representation based on SKL div} and Proposition~\ref{Prop: upper on Gibbs based on KL}, we have
\begin{align}
      &\frac{(1+C_K)D(P_{W|S}^{\gamma}\|P_{W}^{\gamma,L_\mathrm{p}}|P_{S})}{\gamma} \nn\\
    & \leq\genb(P_{{W}|S}^\gamma,P_{S})  \nn\\ 
    & = \frac{D_{\mathrm{SKL}}(P_{W|S}^{\gamma}\|P_{W}^{\gamma,L_\mathrm{p}}|P_{S})}{\gamma} \nn\\
      &=\frac{D(P_{W|S}^{\gamma}\|P_{W}^{\gamma,L_\mathrm{p}}|P_{S})+D(P_{W}^{\gamma,L_\mathrm{p}}\|P_{W|S}^{\gamma}|P_{S})}{\gamma} \nn\\
      &\leq \sqrt{\frac{2\sigma^2 D(P_{W|S}^{\gamma}\|P_{W}^{\gamma,L_\mathrm{p}}|P_{S})}{n}}.
\end{align}
Then, the following upper bound holds for $D(P_{W|S}^{\gamma}\|P_{W}^{\gamma,L_\mathrm{p}}|P_{S})$ 
\begin{align}\label{Eq: kl upper}
    D(P_{W|S}^{\gamma}\|P_{W}^{\gamma,L_\mathrm{p}}|P_{S})\leq \frac{2\sigma^2\gamma^2}{n(1+C_K)^2}.
\end{align}
Using \eqref{Eq: kl upper} in Proposition~\ref{Prop: upper on Gibbs based on KL}, the final result holds.
\end{IEEEproof}

\begin{IEEEproof}[Proof of Theorem~\ref{Theorem: param via conditional SYI}]
From Theorem~\ref{thm:CMI}, and the definition of  $C_C$, we have
\begin{align}
    \genb(P_{{W}|S}^{\gamma},P_S) &= \frac{2 I_{\mathrm{SKL}}(W;U|\widetilde{S})}{\gamma} \nn \\
    & \ge  \frac{2 (1+C_C)  I(W;U|\widetilde{S})}{\gamma}.
\end{align}

Combining with \cite[Theorem 1.2]{steinke2020reasoning}, which states that
\begin{align}
    \genb(P_{{W}|S}^{\gamma},P_S) & \le   \sqrt{\frac{2I(W;U|\widetilde{S})}{n}},
\end{align}
we have
\begin{align}
   \frac{2 (1+C_C)  I(W;U|\widetilde{S})}{\gamma} \le\sqrt{\frac{2I(W;U|\widetilde{S})}{n}},
\end{align}
which gives
\begin{align}
    \genb(P_{{W}|S}^{\gamma},P_S) \leq \frac{\gamma}{(1+C_C)n}.
\end{align}

\end{IEEEproof}

\begin{IEEEproof}[Proof of Theorem~\ref{Theorem: param upper on Gibbs using stability}]
Using the sub-Gaussianity assumption under the Gibbs algorithm and \cite[Lemma~1]{xu2017information}, the following inequality holds,
\begin{align}\label{Eq: bound2}
    &\mathbb{E}_{P_{S,Z}}\left[\mathbb{E}_{P_{W|S^{(i)}}}[\ell(W,Z_i)]-\mathbb{E}_{P_{W|S}}[\ell(W,Z_i)]\right]\\\nonumber&
    \leq \sqrt{2\tau^2 D(P_{W|S^{(i)}}\|P_{W|S}|P_{S,Z})},
\end{align}
Plugging \eqref{Eq: bound2} into \eqref{eq: gen stability rep}, the following upper bound on expected generalization error holds:
\begin{align}\label{Eq: upper gen cond}
    \genb(P_{{W}|S}^{\gamma},P_S) \leq \frac{\sum_{i=1}^n \sqrt{2\tau^2 D(P_{W|S^{(i)}}\|P_{W|S}|P_{S,Z})}}{n}
\end{align}
Comparing \eqref{Eq: upper gen cond} with Theorem~\ref{theorem: Another Representation based on SKL div and stability}, we have:
\begin{align}
    &\frac{(1+C_S)\sum_{i=1}^n
    D(P_{W|S^{(i)}}^{\gamma}\|P_{W|S}^{\gamma}|P_{S,Z})
    }{2\gamma}\nn \\
    &\leq\frac{\sum_{i=1}^n D_{\mathrm{SKL}}(P_{W|S}^{\gamma}\|P_{W|S^{(i)}}^{\gamma}|P_{S,Z})}{2\gamma}\nn\\
    & \leq\frac{\sum_{i=1}^n \sqrt{2\tau^2 D(P_{W|S^{(i)}}\|P_{W|S}|P_{S,Z})}}{n}\nn\\\label{Eq: test 2}
    & \leq\sqrt{\frac{\sum_{i=1}^n 2\tau^2 D(P_{W|S^{(i)}}\|P_{W|S}|P_{S,Z})}{n}},
\end{align}
And we can derive the following upper bound on $\sum_{i=1}^n \sqrt{D(P_{W|S^{(i)}}\|P_{W|S}|P_{S,Z})}$, as follows:
\begin{align}\label{Eq: upper bound 1}
     \sqrt{\sum_{i=1}^n D(P_{W|S^{(i)}}\|P_{W|S}|P_{S,Z})}\leq \frac{2\gamma\sqrt{2\tau^2}}{\sqrt{n}(1+C_S)}
\end{align}
The final result holds by substituting the upper bound in \eqref{Eq: upper bound 1} with \eqref{Eq: test 2}.
\end{IEEEproof}
}
\subsection{Other Tail Distributions}\label{app: More result expected}

\begin{corollary}\label{Corollary: Parametric upper bound using Sub-Exponential}
Suppose that the training samples $S=\{Z_i\}_{i=1}^n$ are i.i.d generated from the distribution $P_Z$, and the non-negative loss function $\ell(w,Z)$ is $(\sigma_e^2,b)$-sub-Exponential on the left-tail \footnote{A random variable $X$ is $(\sigma_e^2,b)$-sub-Exponential on the left-tail if $\log \mathbb{E}[e^{\lambda(X-\mathbb{E}X)}] \le \frac{\sigma_e^2\lambda^2}{2}$, $ -\frac{1}{b}\leq \lambda \leq 0 $.} 
under distribution $P_Z$ for all $w\in \mathcal{W}$. If we further assume $C_I\le \frac{L(W;S)}{I(W;S)}$ for some $C_I \ge 0$, then for the $(\gamma,\pi(w),L_\mathrm{e}(w,s))$-Gibbs algorithm, we have

\begin{align*}
   &\genb(P_{{W}|S}^{\gamma},P_S) \nn\\
   &\ \leq \begin{cases}
    \frac{2\sigma_e^2 \gamma}{n (1+C_I)}, & n\geq \frac{2bI(W;S)}{\sigma_e^2}; \\
\frac{\sigma_e^2}{2b}\Bigl(\frac{\gamma b }{(n(1+C_I)-\gamma b) }+1\Bigr), &  \ceil{\frac{\gamma b}{1+C_I}} <n < \frac{2bI(W;S)}{\sigma_e^2}.
    \end{cases}
\end{align*}
\end{corollary}

\begin{IEEEproof}[Proof of Corollary~\ref{Corollary: Parametric upper bound using Sub-Exponential}]
If the loss function is sub-Exponential on the left-tail we have
$$
\psi^{\star -1}(y)=
\begin{cases}                     &\sqrt{2\sigma_e^2y}, \quad y\leq \frac{\sigma_e^2}{2b};\\
&by+\frac{\sigma_e^2}{2b},\quad \textit{otherwise.}
\end{cases}
$$
If $\frac{I(W;S)}{n}\leq \frac{\sigma_e^2}{2b}$, by Theorem~\ref{Theorem: general psi gibs upper bound}, we have
\begin{align}
    \frac{(1+C_I)I(W;S)}{\gamma}\leq \sqrt{2\sigma_e^2\frac{I(W;S)}{n}},
\end{align}
then the following upper bound holds,
\begin{align}
    I(W;S)\leq \frac{2\sigma_e^2 \gamma^2}{(1+C_I)^2n},
\end{align}
which  gives
\begin{equation}
    \genb(P_{{W}|S}^{\gamma},P_S) \leq \frac{2\sigma_e^2 \gamma}{n (1+C_I)}.
\end{equation}

If $\frac{I(W;S)}{n}> \frac{\sigma_e^2}{2b}$, we have 
\begin{align}
    \frac{I(W;S)(1+C_I)}{\gamma}\leq \frac{b I(W;S)}{n}+\frac{\sigma_e^2}{2b},
\end{align}
then the following upper bound holds when $n> \frac{\gamma b}{1+C_I}$,
\begin{align}
    I(W;S)\leq \frac{\gamma n \sigma_e^2}{2b(n(1+C_I)-\gamma b)},
\end{align}
which gives
\begin{equation*}
    \genb(P_{{W}|S}^{\gamma},P_S) \leq\frac{\sigma_e^2}{2b}\Bigl(\frac{\gamma b }{(n(1+C_I)-\gamma b) }+1 \Bigr). 
\end{equation*}
\end{IEEEproof}

\begin{corollary}\label{Corollary: parametric upper bound using sub-Gamma}
Suppose that the training samples $S=\{Z_i\}_{i=1}^n$ are i.i.d generated from the distribution $P_Z$, and the non-negative loss function $\ell(w,Z)$ is $\Gamma(\tau^2,c_s)$-sub-Gamma on the left-tail \footnote{A random variable $X$ is $\Gamma(\tau^2,c_s)$-sub-Gamma on the left-tail if $\log \mathbb{E}[e^{\lambda(X-\mathbb{E}X)}] \le \frac{\lambda^2 \tau^2}{2(1-c|\lambda|)}$, for $ -\frac{1}{c_s}< \lambda < 0$.}
under distribution $P_Z$ for all $w\in \mathcal{W}$. If we further assume $C_I\le \frac{L(W;S)}{I(W;S)}$ for some $C_I \ge 0$, then for the $(\gamma,\pi(w),L_\mathrm{e}(w,s))$-Gibbs algorithm, if $n>\frac{c_s\gamma}{(1+C_I)}$, we have
\begin{equation}
    \genb(P_{{W}|S}^{\gamma},P_S) \leq \frac{2\tau^2 \gamma}{(1+C_I)n-\gamma c_s}
    \Bigl(1+\frac{\gamma c_s}{(1+C_I)n-\gamma c_s}\Bigr).
\end{equation}
\end{corollary}

\begin{IEEEproof}[Proof of Corollary~\ref{Corollary: parametric upper bound using sub-Gamma}]
By considering $\psi^{\star -1}(y)=\sqrt{2\tau^2y}+c_s y$ in Theorem~\ref{Theorem: general psi gibs upper bound}, we have
\begin{align}
    \frac{(1+C_I)I(W;S)}{\gamma}\leq \sqrt{2\tau^2\frac{I(W;S)}{n}}+c_s\frac{I(W;S)}{n}.
\end{align}
Then the following upper bound holds when $n>\frac{c_s\gamma}{(1+C_I)}$,
\begin{align}
    I(W;S)\leq \Bigl(\frac{\gamma }{(1+C_I)n-\gamma c_s}\Bigr)^2 2n\tau^2,
\end{align}
which gives
\begin{equation*}
    \genb(P_{{W}|S}^{\gamma},P_S) \leq \frac{2\tau^2 \gamma(1+C_I)n}{\Bigl((1+C_I)n-\gamma c_s\Bigr)^2}
    . 
\end{equation*}
\end{IEEEproof}

The authors in \cite{ohnishi2021novel,russo2019much} consider the sub-Exponential assumption for general learning algorithms and provide PAC-Bayesian upper bounds. Similarly, the sub-Gamma assumption is considered in \cite{abou2019exponential,germain2016pac} and PAC-Bayesian upper bounds are provided. 
Our Corollary~\ref{Corollary: Parametric upper bound using Sub-Exponential} and~\ref{Corollary: parametric upper bound using sub-Gamma}  provide upper bounds with order $\mathcal{O}(1/n)$ on the expected generalization error for Gibbs algorithm under these assumptions.


\section{PAC-Bayesian Upper Bound}\label{app: PAC-Bayesian upper bound} 
Since the $(\gamma,\pi(w),L_\mathrm{p}(w,P_{S^\prime}))$-Gibbs distribution only depends on the population risk $L_\mathrm{p}(w,P_{S^\prime})$ and is independent of the samples $S$, we can denote it as $P_W^{\gamma,L_\mathrm{p}^\prime}$.

\begin{IEEEproof}[Proof of Theorem~\ref{Theorem: PAC Baysian result}]
Using Lemma~\ref{Lemma: exact SKL two Gibbs}, we have
\begin{align}\label{equ:PAC_main}
&D_{\mathrm{SKL}}(P_{W|S}^{\gamma}\|P_{W}^{\gamma,L_\mathrm{p}^\prime})\nn\\
&\ =\gamma(\mathbb{E}_{P_{W|S=s}^\gamma}[L_\mathrm{p}(W,P_{Z^\prime})]- \mathbb{E}_{P_{W|S=s}^\gamma}[L_\mathrm{e}(W,s)])\nn\\
&\qquad{}-\gamma(\mathbb{E}_{P_{W}^{\gamma,L_\mathrm{p}^\prime}}[L_\mathrm{p}(W,P_{Z^\prime})]- \mathbb{E}_{P_{W}^{\gamma,L_\mathrm{p}^\prime}}[L_\mathrm{e}(W,s)])\nn\\
&\ \leq \gamma\left|\mathbb{E}_{P_{W|S=s}^\gamma}[L_\mathrm{p}(W,P_{Z^\prime})]- \mathbb{E}_{P_{W|S=s}^\gamma}[L_\mathrm{e}(W,s)]\right|\nn\\
&\qquad{}+\gamma\left|(\mathbb{E}_{P_{W}^{\gamma,L_\mathrm{p}^\prime}}[L_\mathrm{p}(W,P_{Z^\prime})]- \mathbb{E}_{P_{W}^{\gamma,L_\mathrm{p}^\prime}}[L_\mathrm{e}(W,s)]\right|
\nn\\
&\ \leq \gamma\left|\mathbb{E}_{P_{W|S=s}^{\gamma}}[L_\mathrm{p}(W,P_{Z^\prime})]- \mathbb{E}_{P_{W|S=s}^{\gamma}}[L_\mathrm{p}(W,P_Z)]\right| 
\nn\\
&\qquad{}+\gamma\left|\mathbb{E}_{P_{W|S=s}^\gamma}[L_\mathrm{p}(W,P_Z)]- \mathbb{E}_{P_{W|S=s}^\gamma}[L_\mathrm{e}(W,s)]\right|
\nn\\
&\qquad\quad{}+\gamma\left|\mathbb{E}_{P_{W}^{\gamma,L_\mathrm{p}^\prime}}[L_\mathrm{p}(W,P_{Z^\prime})]- \mathbb{E}_{P_{W}^{\gamma,L_\mathrm{p}^\prime}}[L_\mathrm{p}(W,P_Z)]\right|
\nn\\
&\qquad\qquad{}+\gamma\left|\mathbb{E}_{P_{W}^{\gamma,L_\mathrm{p}^\prime}}[L_\mathrm{p}(W,P_Z)]- \mathbb{E}_{P_{W}^{\gamma,L_\mathrm{p}^\prime}}[L_\mathrm{e}(W,s)]\right|,
\end{align}
and we just need to bound the four terms in the above inequality.

The first and the third term in~\eqref{equ:PAC_main} can be bounded using the Donsker-Varadhan variational characterization of KL divergence, note that for all $\lambda \in \mathbb{R}$,
\begin{align}
 D(P_{Z^\prime}\| P_Z) &\ge \mathbb{E}_{P_{Z^\prime}}[\lambda \ell(w,Z^\prime)] - \log \mathbb{E}_{P_{Z}}[e^{\lambda \ell(w,Z)}]\nn \\
 & \ge \lambda (L_\mathrm{p}(w,P_{Z^\prime})-L_\mathrm{p}(w,P_Z))-\frac{\lambda^2 \sigma^2}{2},
\end{align}
where the last step follows from the sub-Gaussian assumption. Since the above inequality holds for all $\lambda \in \mathbb{R}$, the discriminant must be non-positive, which
implies
\begin{align}\label{eq: PAC1}
    &|L_\mathrm{p}(w,P_{Z^\prime})-L_\mathrm{p}(w,P_Z)|\leq \sqrt{2\sigma^2 D(P_{Z^\prime}\| P_Z)}, \quad w\in \mathcal{W}.
\end{align}

We use the PAC-Bayesian bound in \cite[Proposition 3]{guedj2019still} to bound the second and the fourth term in ~\eqref{equ:PAC_main}. For any posterior distribution $Q_{W|S=s}$, and prior distribution $Q_W$, if $\ell(w,Z)$ is $\sigma$-sub-Gaussian under $P_Z$ for all $w \in \mathcal{W}$, the following bound holds with probability $1-\delta$,
\begin{align*}
   &\left|\mathbb{E}_{Q_{W|S=s}}[L_\mathrm{p}(W,P_Z)]- \mathbb{E}_{Q_{W|S=s}}[L_\mathrm{e}(W,s)]\right|\\
   &\qquad\qquad\qquad\qquad\leq \sqrt{\frac{2\sigma^2\Bigl( D(Q_{W|S=s}\|Q_{W})+\log(1/\delta)\Bigr)}{n}}.
\end{align*}

If we choose $P_{W|S}^{\gamma}$ as the posterior distribution and $P_{W}^{\gamma,L_\mathrm{p}^\prime}$ as the prior distribution, we have
\begin{align}\label{eq: PAC2}
   &\left|\mathbb{E}_{P_{W|S=s}^\gamma}[L_\mathrm{p}(W,P_Z)]- \mathbb{E}_{P_{W|S=s}^\gamma}[L_\mathrm{e}(W,s)]\right|\nn\\
   &\qquad\qquad\qquad\quad\leq \sqrt{\frac{2\sigma^2\left( D(P_{{W}|S=s}^{\gamma}\|P_{W}^{\gamma,L_\mathrm{p}^\prime})+\log(1/\delta)\right)}{n}}
\end{align}
holds with probability $1-\delta$.
If we set $Q_{W|S=s} = Q_{W} = P_{W}^{\gamma,L_\mathrm{p}^\prime}$, we have
\begin{align}\label{eq: PAC4}
   &\left|\mathbb{E}_{P_{W}^{\gamma,L_\mathrm{p}^\prime}}[L_\mathrm{p}(W,P_{Z})]- \mathbb{E}_{P_{W}^{\gamma,L_\mathrm{p}^\prime}}[L_\mathrm{e}(W,s)]\right|\nn\\
   &\qquad\qquad\qquad\qquad\qquad\qquad\leq \sqrt{\frac{2\sigma^2\left( \log(1/\delta)\right)}{n}}.
\end{align}

Combining the bounds in \eqref{eq: PAC1}, \eqref{eq: PAC2} and \eqref{eq: PAC4} with \eqref{equ:PAC_main}, we have
\begin{align}
&D_{\mathrm{SKL}}(P_{W|S}^{\gamma}\|P_{W}^{\gamma,L_\mathrm{p}^\prime})\nn\\
&\ \leq \gamma\sqrt{\frac{2\sigma^2\left( D(P_{{W}|S=s}^{\gamma}\|P_{W|S}^{\gamma,L_\mathrm{p}^\prime})+\log(1/\delta)\right)}{n}}\nn\\
&\qquad\qquad\qquad{}+\gamma \sqrt{\frac{2\sigma^2\left( \log(1/\delta)\right)}{n}}
+2\gamma \sqrt{2\sigma^2 D(P_{Z^\prime}\| P_Z)}.
\end{align}
Then,  using the assumption that 
\begin{equation*}
(1+C_P(s))D(P_{W|S=s}^{\gamma}\|P_{W}^{\gamma,L_\mathrm{p}^\prime})\le D_{\mathrm{SKL}}(P_{W|S}^{\gamma}\|P_{W|S}^{\gamma,L_\mathrm{p}}),
\end{equation*}
we have
\begin{align}
&(1+C_P(s))D(P_{W|S=s}^{\gamma}\|P_{W}^{\gamma,L_\mathrm{p}^\prime})\nn\\
&\qquad \leq \gamma\sqrt{\frac{2\sigma^2\left( D(P_{{W}|S=s}^{\gamma}\|P_{W}^{\gamma,L_\mathrm{p}^\prime})+\log(1/\delta)\right)}{n}}\nn\\
&\qquad\qquad{}+\gamma \sqrt{\frac{2\sigma^2\left( \log(1/\delta)\right)}{n}}
+2\gamma \sqrt{2\sigma^2 D(P_{Z^\prime}\|P_Z)}.
\end{align}
Denote $\gamma^\prime\triangleq\gamma/(1+C_P(s))$, then we have
\begin{align}\label{eq: PAC 5}
&D(P_{W|S=s}^{\gamma}\|P_{W}^{\gamma,L_\mathrm{p}^\prime})\nn\\
&\qquad{}- \sqrt{\frac{2{\gamma^\prime}^2\sigma^2\left( \log(1/\delta)\right)}{n}}
- \sqrt{8 {\gamma^\prime}^2\sigma^2 D(P_{Z^\prime}\|P_Z)}\nn\\
&\qquad\qquad\leq \sqrt{\frac{2{\gamma^\prime}^2\sigma^2\left( D(P_{{W}|S=s}^{\gamma}\|P_{W}^{\gamma,L_\mathrm{p}^\prime})+\log(1/\delta)\right)}{n}}.
\end{align}
If we have 
\begin{align*}
0 &\leq D(P_{W|S=s}^{\gamma}\|P_{W}^{\gamma,L_\mathrm{p}^\prime})\nn\\
&\leq \sqrt{\frac{2{\gamma^\prime}^2\sigma^2\left( \log(1/\delta)\right)}{n}}
+ \sqrt{8 {\gamma^\prime}^2\sigma^2 D(P_{Z^\prime}\|P_Z)},
\end{align*}
then the above inequality holds. Otherwise, we could take square over both sides in \eqref{eq: PAC 5}, 
and denote 
\begin{equation*}
   A\triangleq C+\sqrt{\frac{2\sigma^2{\gamma^\prime}^2\log(1/\delta)}{n}},\quad  B\triangleq \sqrt{8{\gamma^\prime}^2 \sigma^2 D(P_{Z^\prime}\|P_Z)},
\end{equation*}
where $C\triangleq\sigma^2 {\gamma^\prime}^2/n$, then we have
\begin{align}\label{eq: PAC 6}
D^2(P_{W|S=s}^{\gamma}\|P_{W}^{\gamma,L_\mathrm{p}^\prime})
 &-2D(P_{W|S=s}^{\gamma}\|P_{W}^{\gamma,L_\mathrm{p}^\prime})(A+B)\nn\\
 &\qquad\qquad{}+B^2+2(A-C)B\leq 0.
\end{align}
Solving the above inequality gives:
\begin{equation}
     0\le D(P_{W|S=s}^{\gamma}\|P_{W}^{\gamma,L_\mathrm{p}^\prime})\leq \sqrt{A^2+2BC}+A+B.
\end{equation}
As $\sqrt{x+y}\leq \sqrt{x}+\sqrt{y}$ for positive $x,y$ and $A\ge C$, we have
\begin{align}\label{eq: PAC 7}
   D(P_{W|S=s}^{\gamma}\|P_{W}^{\gamma,L_\mathrm{p}^\prime})
   &\leq 2A+B+\sqrt{2BC}\nn\\
   &\leq 2A+B+\sqrt{2AB}\nn\\
   &\leq (\sqrt{2A}+\sqrt{B})^2.
\end{align}

Now using \eqref{eq: PAC 7} in \eqref{eq: PAC2} and applying the inequality $\sqrt{x+y}\leq \sqrt{x}+\sqrt{y}$, we have
\begin{align*}\label{Eq: Final PAC Union}
  &  \left|\mathbb{E}_{P_{W|S=s}^\gamma}[L_\mathrm{p}(W,\mu)-L_\mathrm{e}(W,s)]\right| \\
  &\quad \leq \sqrt{\frac{2\sigma^2(\sqrt{2A}+\sqrt{B})^2+2\sigma^2\log(1/\delta)}{n}}\\
    &\quad\leq \sqrt{\frac{4\sigma^2A}{n}}+\sqrt{\frac{2\sigma^2 B}{n}}+\sqrt{\frac{2\sigma^2\log(1/\delta)}{n}}
    \\
    &\quad\leq\frac{2\gamma\sigma^2}{(1+C_P(s))n} + \sqrt{\frac{2\sigma^2\left( \log(1/\delta)\right)}{n}}
    \\
    &\quad\qquad{} +2\sqrt{\frac{\gamma\sigma^2}{(1+C_P(s))n}}\nn\\
    &\quad\qquad\qquad\qquad{}\cdot\left(\sqrt[4]{\frac{2\sigma^2\log(1/\delta)}{n}}+\sqrt[4]{2\sigma^2 D(P_{Z^\prime}\|P_Z)}\right).
\end{align*}
As both \eqref{eq: PAC2} and \eqref{eq: PAC4}  hold with probability at least $1-\delta$, the above inequality holds with probability at least $1-2\delta$ by the union bound \cite{vershynin2018high}.
\end{IEEEproof}





\section{Asymptotic Behavior of Generalization Error for Gibbs Algorithm}

\subsection{Large Inverse Temperature Details}\label{app: large alpha}

\begin{IEEEproof}[Proof of Proposition~\ref{prop:single_well}]
It is shown in \cite{athreya2010gibbs,hwang1980laplace} that if the following Hessian matrix
\begin{equation}
        H^*(S) = \nabla^2_w L_\mathrm{e} (w,S)\big|_{w = W^*(S)}
\end{equation}
is not singular, then as $\gamma \to \infty$
\begin{equation}
   P_{W|S}^\gamma \to \mathcal{N}(W^*(S), \frac{1}{\gamma}H^*(S)^{-1}  )
\end{equation}
in distribution.
Then, the mean of the marginal distribution $P_W$ equals to the mean of $W^*(S)$, i.e.,
\begin{equation}
    \mathbb{E}_{P_W}[W] = \mathbb{E}_{P_S}[W^*(S)].
\end{equation}
To apply Theorem~\ref{Theorem: Gibbs Result}, we evaluate the symmetrized KL information using the Gaussian approximation:
\begin{align}
    & I(W;S)+L(W;S) \nn \\
    & = \mathbb{E}_{P_{W,S}}[\log P_{W|S}^{\gamma}] -  \mathbb{E}_{P_{W}\otimes P_{S}}[\log P_{W|S}^{\gamma}]\nn \\
    & = \mathbb{E}_{P_{W,S}}\Bigl[-\frac{\gamma}{2}(W - W^*(S))^\top H^*(S) (W - W^*(S))\Bigr] \nn\\ &\qquad+\mathbb{E}_{P_{W}\otimes P_{S}}\Bigl[\frac{\gamma}{2}(W - W^*(S))^\top H^*(S) (W - W^*(S))\Bigr] \nn\\
    & = \mathbb{E}_{P_{W}\otimes P_{S}}\Bigl[\frac{\gamma}{2}W^\top H^*(S) W\Bigr]-\mathbb{E}_{P_{W,S}}\Bigl[\frac{\gamma}{2}W^\top H^*(S)W\Bigr]\nn\\
    &\qquad + \mathbb{E}_{P_{S}\otimes P_{W}}\Bigl[\frac{\gamma}{2}\Bigl(\tr\Bigl(H^*(S) (W^*(S)W^*(S)^\top\nn\\&\qquad\quad - WW^*(S)^\top-W^*(S)W^\top)\Bigr) \Bigr) \Bigr]\nn\\
    &\qquad - \mathbb{E}_{P_{S}\otimes P_{W|S}}\Bigl[\frac{\gamma}{2}\Bigl(\tr\Bigl(H^*(S) (W^*(S)W^*(S)^\top\nn\\&\qquad\quad - WW^*(S)^\top-W^*(S)W^\top)\Bigr) \Bigr) \Bigr].
\end{align}
Note that $\mathbb{E}_{P_{W}}[W] =\mathbb{E}_{P_S}[W^*(S)]$ and  $\mathbb{E}_{P_{W|S}}[W] =W^*(S)$, we have
\begin{align*}
    &\genb(P_{{W}|S}^\infty,\mu)\nn\\
    &\ =
    \frac{I(W;S)+L(W;S)}{\gamma}\\
    &\ = \mathbb{E}_{P_{W}\otimes P_{S}}\Bigl[\frac{1}{2}W^\top H^*(S) W\Bigr]-\mathbb{E}_{P_{W,S}}\Bigl[\frac{1}{2}W^\top H^*(S)W\Bigr]\\
    &\qquad{} + \mathbb{E}_{ P_{S}}\Bigl[\frac{1}{2}\Bigl(\tr\Bigl(H^*(S) \Bigl( -  \mathbb{E}[W^*(S)]\,W^*(S)^\top\nn\\
    &\quad\qquad\qquad\qquad\qquad\qquad\qquad{}-W^*(S)\,\mathbb{E}[W^*(S)]^\top\Bigr)\Bigr) \Bigr) \Bigr]\\
    &\qquad{} - \mathbb{E}_{P_{S}}\Bigl[\frac{1}{2}\Bigl(\tr\Bigl(H^*(S) \Bigl( - W^*(S)W^*(S)^\top\nn\\
    &\quad\qquad\qquad\qquad\qquad\qquad\qquad{}-W^*(S)W^*(S)^\top\Bigr)\Bigr) \Bigr) \Bigr]\\
    & = \mathbb{E}_{P_{W}\otimes P_{S}}\Bigl[\frac{1}{2}W^\top H^*(S) W\Bigr]-\mathbb{E}_{P_{W,S}}\Bigl[\frac{1}{2}W^\top H^*(S)W\Bigr]\\
    &\qquad{} +\mathbb{E}_{P_{S}}\Bigl[ (W^*(S)- \mathbb{E}[W^*(S)])^\top \nn\\
    &\qquad\qquad\qquad\qquad{}\cdot \bigl(H^*(S)\,W^*(S)- \mathbb{E}[H^*(S)\,W^*(S)]\bigr) \Bigr]. 
\end{align*}
\end{IEEEproof}

\begin{IEEEproof}[Proof of Proposition~\ref{prop:multiple_well}]
In this multiple-well case, it is shown in \cite{athreya2010gibbs} that the Gibbs algorithm can be approximated by the following Gaussian mixture distribution
\begin{equation}
  P_{W|S}^\gamma \to \frac{\sum_{u=1}^M  \pi\Bigl(W_u^*(S)\Bigr)\mathcal{N}\Bigl(W_u^*(S), \frac{1}{\gamma}H^*_u(S)^{-1} \Bigr)}{\sum_{u=1}^M \pi(W_u^*(S))} ,
\end{equation}
as long as $H^*_u(S) \triangleq \nabla^2_w L_\mathrm{e} (w,S)\big|_{w = W_u^*(S)}$ is not singular for all $u\in \{1,\cdots, M\}$.

However, there is no closed form for the symmetrized KL information for Gaussian mixtures. Thus,  we use Theorem~\ref{Theorem: Gibbs Result} to construct an upper bound of the generalization error.

Consider the latent random variable $U\in \{1,\cdots,M\}$ which denotes the index of the Gaussian component of $P_{W|S}^\gamma$. Then, conditioning on $U$ and $S$, $W$ is a Gaussian random variable. Moreover, since $\pi(W)$ is a uniform prior, $U$ is a discrete uniform distribution $P_U(U=u)=\frac{1}{M}$, and $U\perp S$. Note that for mutual information, we have
\begin{align*}
    I(S;W|U) &= I(S;W|U) +I(S;U)\\
    &= I(S;W,U)\\
    &= I(S;W) + I(S;U|W) \\
    &\ge I(S;W),
\end{align*}
and for lautum information
\begin{align*}
    L(W;S) &\overset{(a)}{\le} L(W,U;S)\\ &\overset{(b)}{=} L(U;S) + L(W;S|U)\\ &= L(W;S|U),
\end{align*}
where $(a)$ is due to the data processing inequality for any $f$-divergence, and $(b)$ follows by the fact that the chain rule of  lautum information holds when $U\perp S$ as shown in \cite{palomar2008lautum}.

Thus, we can upper bound $I(S;W)$ and $L(S;W)$ with $I(S;W|U)$ and $L(S;W|U)$, respectively,
\begin{align*}
    &\genb(P_{{W}|S}^\infty,\mu) \\
    &\ = \lim_{\gamma \to \infty }\frac{I(S;W)+L(S;W)}{\gamma} \\
    &\ \le \lim_{\gamma \to \infty }\frac {I(S;W|U) + L(S;W|U)}{\gamma}\\
    &\ = \mathbb{E}_U \Bigl[\mathbb{E}_{P_{W|U}\otimes P_{S}}\Bigl[\frac{1}{2}W^\top H(w_u^*(S),S) W\Bigr]\Bigr]\nn\\
    &\qquad{}-\mathbb{E}_{U}\Bigl[\mathbb{E}_{P_{W,S|U}}\Bigl[\frac{1}{2}W^\top H(w_U^*(S),S)W\Bigr]\Bigr]\\
    &\qquad\qquad{} +\mathbb{E}_{U}\Bigl[\mathbb{E}_{P_{S}}\Bigl[ (w_U^*(S) - \mathbb{E}[w_U^*(S)])^\top\nn\\ &\qquad\qquad\qquad{}\cdot \bigl(H(w_U^*(S),S)w_U^*(S)\nn\\
    &\qquad\qquad\qquad\qquad{}- \mathbb{E}[H(w_U^*(S),S)w_U^*(S)]\bigr) \Bigr] \Bigr]. 
\end{align*}
\end{IEEEproof}

\subsection{Regularity Conditions for MLE}\label{app:MLE}

In this section, we present the regularity conditions required by the asymptotic normality \cite{van2000asymptotic} of maximum likelihood estimates. 

\begin{assumption}\label{assump:MLE}
\textbf{Regularity Conditions for MLE:}
\begin{enumerate}
  \item $f(z|\vw) \ne f(z|\vw')$ for $\vw \ne \vw'$.
  \item $\mathcal{W}$ is an open subset of $\mathbb{R}^d$.
  \item The function $\log f(z|\vw)$ is three times continuously differentiable with
respect to $\vw$.
  \item There exist functions $F_1(z)\colon\mathcal{Z} \to \mathbb{R}$, $F_2(z)\colon\mathcal{Z} \to \mathbb{R}$ and $M(z)\colon\mathcal{Z} \to \mathbb{R}$, such that
\begin{equation*}
  \mathbb{E}_{Z\sim f(z|\vw)}[M(Z)] <\infty,
\end{equation*}
and the following inequalities hold for any $\vw \in \mathcal{W}$:
\begin{align*}
  \left|\frac{\partial \log f(z|\vw)}{\partial w_i} \right|<F_1(z), &\quad  \left|\frac{\partial^2 \log f(z|\vw)}{\partial w_i \partial w_j} \right|<F_2(z), \\
  \left|\frac{\partial^3 \log f(z|\vw)}{\partial w_i \partial w_j \partial w_k} \right|<M(z), &\qquad i,j,k =1,2,\cdots,d.
\end{align*}
  \item The following inequality holds for an arbitrary $\vw \in \mathcal{W}$ and $i,j=1,2,\cdots,d$:
\begin{equation*}
  0< \mathbb{E}_{Z\sim f(z|\vw)}\left[\frac{\partial \log f(z|\vw)}{\partial w_i}\frac{\partial \log f(z|\vw)}{\partial w_j}\right] <\infty.
\end{equation*}
\end{enumerate}
\end{assumption}

\subsection{Bayesian Learning Algorithm}\label{app:bayesian_learning}
In this section, we show that the symmetrized KL information can be used to characterize the generalization error of Gibbs algorithm in a different asymptotic regime, i.e., inverse temperature $\gamma=n$, then  $\gamma$ and $n$ go to infinity simultaneously. In this regime, the Gibbs algorithm is equivalent to the  Bayesian posterior distribution instead of ERM.

Suppose that we have $n$ i.i.d. training samples $S = \{Z_i\}_{i=1}^n$ generated from the distribution $P_Z$ defined on $\mathcal{Z}$, and we want to fit the training data with a parametric distribution family $\{f(z_i|\vw)\}_{i=1}^n$, where $\vw\in \mathcal{W}\subset \mathbb{R}^d$ denotes the parameter and $\pi({\vw})$ denotes a pre-selected prior distribution. Here, the true data-generating distribution may not belong to the parametric family, i.e., $P_Z \ne f(\cdot|\vw)$ for $\vw \in \mathcal{W}$.  The following Bayesian posterior distribution
\begin{subequations}
\begin{equation}
    P_{{W}|S}(\vw|z^n) = \frac{\pi({\vw})\prod_{i}^n f(z_i|{\vw})}{V(z^n)}
\end{equation}
with
\begin{equation}
V(z^n) = \int \pi(\vw) \prod_{i}^n f(z_i|\vw)\, \diff \vw,
\end{equation}
\label{equ:posterior}%
\end{subequations}
is equivalent to the $(n,\pi(\vw), L_\mathrm{e}(\vw,s))$-Gibbs algorithm with log-loss $\ell(\vw,z)=-\log f(z|\vw)$.
Thus, Theorem \ref{Theorem: Gibbs Result} can be applied directly, and we just need to evaluate $I_{\mathrm{SKL}}(W;S)$.





We further assume that the parametric family $\{f(z|\vw), \vw \in \mathcal{W}\}$ and prior $\pi(\vw)$ satisfy all the regularization conditions required for the Bernstein–von-Mises theorem \cite{van2000asymptotic} and the asymptotic Normality of the maximum likelihood estimate (MLE), including Assumption~\ref{assump:MLE} and the condition that $\pi(w)$ is continuous and $\pi(w)>0$ for all $w \in \mathcal{W}$.

In the asymptotic regime $n\to \infty$, Bernstein–von-Mises theorem under model mismatch \cite{van2000asymptotic,kleijn2012bernstein} states that we could approximate the Bayesian posterior distribution $P_{W|S}$ in \eqref{equ:posterior} by
\begin{equation*}
   \mathcal{N}(\hat{W}_{\mathrm{ML}}, \frac{1}{n} J(\vw^*)^{-1}),
\end{equation*}
where
\begin{equation*}
    \hat{W}_{\mathrm{ML}} \triangleq \argmax_{\vw\in \mathcal{W}} \sum_{i=1}^n \log f(Z_i|\vw),
\end{equation*}
denotes the MLE and
\begin{equation*}
    J(\vw) \triangleq \E_Z \Bigl[ -\nabla_{\vw}^2 \log f(Z|\vw) \Bigr]
\end{equation*}
with
\begin{equation*}
\vw^*\triangleq \argmin_{\vw\in \mathcal{W}} D(P_Z\|f(\cdot|\vw)).
\end{equation*}
The asymptotic Normality of the MLE states that  the distribution of $\hat{W}_{\mathrm{ML}}$ will converge to
\begin{equation*}
    \mathcal{N}(\vw^*, \frac{1}{n} J(\vw^*)^{-1} \mathcal{I}(\vw^*)J(\vw^*)^{-1})
\end{equation*}  
with
\begin{equation*}
\mathcal{I}(\vw) \triangleq \E_Z \Bigl[ \nabla_{\vw} \log f(Z|\vw) \nabla_{\vw} \log f(Z|\vw) ^\top\Bigr]
\end{equation*}
as $n \to \infty$. Thus, the marginal distribution $P_{W}$ can be approximated by a Gaussian distribution regardless the choice of prior $\pi(\vw)$.


Then, the symmetrized KL information can be computed using  Lemma~\ref{lemma:Gaussian}. By Theorem~\ref{Theorem: Gibbs Result}, we have
\begin{align}
  \genb(P_{W|S},P_Z)
    = \frac{I_{\mathrm{SKL}}(S;W)}{n} = \frac{\tr(\mathcal{I}(\vw^*)J(\vw^*)^{-1})}{n}.
\end{align}
When the true model is in the parametric family $P_Z = f(\cdot|\vw^*)$, we have $\mathcal{I}(\vw^*)=J(\vw^*)$, which gives the Fisher information matrix and $\genb(P_{W|S},P_Z) =\frac{d}{n}$. This result suggests that the expected generalization error of MLE and that of the Bayesian posterior distribution are the same under suitable regularity conditions.

\subsection{Behavior of Empirical Risk}\label{App: EP behaviuor}
As an aside, we show that the empirical risk is a decreasing function of the inverse temperature $\gamma$.
To see this, we first note that the derivative of $P_{W|S}^{\gamma}$ with respect to $\gamma$ is given by
\begin{align}
    \frac{\diff P_{{W}|S}^\gamma (w|s)}{\diff \gamma} =P_{{W}|S}^\gamma (w|s) \left(\mathbb{E}_{P_{{W}|S}^\gamma}[L_\mathrm{e}(w,S)]-L_\mathrm{e}(w,S)\right).
\end{align}
Then, we can compute the derivative of the empirical risk with respect to $\gamma$ as follows:
\begin{align}
    &\frac{\diff \mathbb{E}_{P_{W,S}}[L_\mathrm{e}(W,S)]}{\diff\gamma}\nn\\
    &\ =\mathbb{E}_{P_S}[\frac{\diff \mathbb{E}_{P_{W|S}^\gamma}[L_\mathrm{e}(W,S)]}{\diff\gamma}]\nn \\
    &\ =\mathbb{E}_{P_S}\left[\int_{\mathcal{W}} L_\mathrm{e}(w,S) \frac{\diff P_{{W}|S}^\gamma (w|S)}{\diff \gamma} dw \right]\nn \\
    &\ =\mathbb{E}_{P_S}\biggl[\int_{\mathcal{W}}  P_{{W}|S}^\gamma (w|s) \Bigl(L_\mathrm{e}(w,S)\, \mathbb{E}_{P_{{W}|S}^\gamma}[L_\mathrm{e}(w,S)]\nn\\
    &\quad\qquad\qquad\qquad\qquad\qquad\qquad\qquad\qquad{}-L_\mathrm{e}^2(w,S)\Bigr)\, \diff w  \biggr]\nn \\
    &\ =\mathbb{E}_{P_S}\left[ \mathbb{E}_{P_{{W}|S}^\gamma}^2[L_\mathrm{e}(w,S)]-\mathbb{E}_{P_{{W}|S}^\gamma}[L_\mathrm{e}^2(w,S)]\right]\nn \\
    &\ =-\mathbb{E}_{P_S}[\textrm{Var}_{P_{{W}|S}^\gamma}[L_\mathrm{e}(W,S)]]\leq 0
\end{align}
When $\gamma=0$, it can be shown that $(0,\pi(w),L_\mathrm{e}(w,s))$-Gibbs algorithm has zero generalization error. However, the empirical risk in this case could be large, since the training samples are not used at all. As $\gamma \to \infty$, the empirical risk is decreasing, but the generalization error could be large. Thus, the inverse temperature $\gamma$ controls the trade-off between the empirical risk and the generalization error.

%

\section{Regularized Gibbs Algorithm}\label{app: Regularizer}
\begin{IEEEproof}[Proof of Proposition~\ref{Prop: Prop Regularizer}]
For $(\gamma,\pi(w),L_\mathrm{e}(w,s) + \lambda R(w,s))$-Gibbs algorithm, we have
\begin{align*}
    I_{\mathrm{SKL}}(W;S)&=\mathbb{E}_{P_{W,S}}[\log(P_{W|S}^\gamma)]-\mathbb{E}_{P_W\otimes P_S}[\log(P_{W|S}^\gamma)]\\
    &=\gamma\left(\mathbb{E}_{P_W\otimes P_S}[L_\mathrm{e}(W,S)] -\mathbb{E}_{P_{W,S}}[L_\mathrm{e}(W,S)]\right)\\\nonumber
    &\quad+\gamma \lambda \left(\mathbb{E}_{P_W\otimes P_S}[R(W,S)]-\mathbb{E}_{P_{W,S}}[R(W,S)]\right)\\
    &=\gamma\genb(P_{{W}|S}^{\gamma},P_S)+\gamma \lambda\mathbb{E}_{\Delta_{W,S}}[R(W,S)]. 
\end{align*}
\end{IEEEproof}

\begin{IEEEproof}[Proof of Corollary~\ref{Corollary: l2 regularizer}]
We just need to compute $\mathbb{E}_{\Delta_{W,S}}[R(W,S)]$ by considering $R(w,s)=\|w-T(s) \|_2^2$,
\begin{align*}
    &\mathbb{E}_{P_W\otimes P_S}[R(W,S)] - \mathbb{E}_{P_{W,S}}[R(W,S)]\\
    &=\mathbb{E}_{P_W\otimes P_S}\left[\|W-T(S) \|_2^2\right] - \mathbb{E}_{P_{W,S}}\left[\|W-T(S) \|_2^2\right]\\
    &=\mathbb{E}_{P_{W,S}}\left[W^TT(S)\right] - \mathbb{E}_{P_W\otimes P_S}\left[W^TT(S)\right]\\
    &=\tr(\Cov(W,T(S))). 
\end{align*}
\end{IEEEproof}
\begin{IEEEproof}[Proof of Proposition~\ref{Proposition: Regularizer general bound}]
Using the decoupling lemma from \cite[Theorem 1]{bu2020tightening}, we have:
\begin{align}
    |\mathbb{E}_{\Delta_{W,S}}[R(W,S)]|\leq\psi^{*-1}( I(W;S)),
\end{align}
which means that
\begin{equation}\label{eq: donsker result}
  -\psi^{*-1}( I(W;S))  \leq \mathbb{E}_{\Delta_{W,S}}[R(W,S)] \leq \psi^{*-1}( I(W;S)). 
\end{equation}
The final results \eqref{eq: reg sub main} follows directly from \eqref{eq: donsker result} and Proposition~\ref{Prop: Prop Regularizer}.
\end{IEEEproof}
\begin{IEEEproof}[Proof of Corollary~\ref{Proposition: Regularizer sub-gaussian}]
Considering $\psi^{*-1}( I(W;S))=\sqrt{2\sigma^2 I(W;S)}$ in Proposition~\ref{Proposition: Regularizer general bound} completes the proof.
\end{IEEEproof}

By assuming $\sigma$-sub-Gaussianity  for both loss function and the regularizer, we provide a generalization error upper bound for regularized Gibbs algorithm in the following proposition.

{\color{black}

}

\end{appendices}

\bibliographystyle{IEEEtran}
\bibliography{Refs}

\end{document}